\documentclass[aps,floats,nofootinbib,amssymb,preprint,superscriptaddress]{revtex4}
\usepackage{epsf,epsfig}
\usepackage{subfigure}
\usepackage{graphicx}
\usepackage{color}

\newcommand{\be}{\begin{equation}}
\newcommand{\ee}{\end{equation}}
\newcommand{\bea}{\begin{eqnarray}}
\newcommand{\eea}{\end{eqnarray}}

\renewcommand{\check}{({\bf This statement needs to be checked!!})}

\newcommand{\half}{\frac{1}{2}}

\newcommand{\ba}{\begin{array}}
\newcommand{\ea}{\end{array}}
\newcommand{\bi}{\begin{itemize}}
\newcommand{\ei}{\end{itemize}}
\newcommand{\ben}{\begin{enumerate}}
\newcommand{\een}{\end{enumerate}}

\newcommand{\cs}{\mathbb S}

\global\long\def\met{\not{\!{\rm E}}_{T}}

\preprint{
\hbox to \hsize{
\hfill$\vcenter{\hbox{\bf MAD-PH-10-1558}
	\hbox{\bf NUHEP-TH/10-04}
	\hbox{\bf ANL-HEP-PR-10-12}
	\hbox{May 2010}}$}
}

\begin{document}
%-----------------------------------
% Title
%-----------------------------------
\title{\vspace*{.75in}
Complex Scalar Dark Matter vis-\`{a}-vis CoGeNT, DAMA/LIBRA and XENON100}
%-----------------------------------
% Authors & Address
%-----------------------------------
\author{Vernon Barger}
\affiliation{Department of Physics, University of Wisconsin, Madison, WI 53706}
\author{Mathew McCaskey}
\affiliation{Department of Physics, University of Wisconsin, Madison, WI 53706}
\author{Gabe Shaughnessy}
\affiliation{Northwestern University, Department of Physics and Astronomy, Evanston, IL 60208 USA}
\affiliation{HEP Division, Argonne National Lab, Argonne IL 60439 USA}

\thispagestyle{empty}

\begin{abstract}
The CoGeNT and DAMA/LIBRA experiments have found evidence for the spin-independent scattering from nuclei of a light dark matter (DM) particle, 7-12 GeV, which is not excluded by the XENON DM experiments.  We show that this putative DM signal can be explained by a complex scalar singlet extension of the standard model (CSM), with a thermal cosmological DM density, and a Higgs sector that is consistent with LEP constraints.  We make predictions for the masses, production, and decays of the two Higgs mass eigenstates and describe how the Higgs and DM particles can be discovered at the LHC.  
\end{abstract}
\date{\today}
\maketitle
%=============================================================
% BEGIN MAIN TEXT
%=============================================================
\newpage

%%%%%%%%%%%%%%%%%%%%%%%%%%%%%%%%%%
\section{Introduction}
\label{sect:intro}
%%%%%%%%%%%%%%%%%%%%%%%%%%%%%%%%%%

There has been a flurry of recent activity concerning possible experimental signals of particle dark matter (DM) and numerous theoretical models have been put forward to explain them.  The standard model (SM) has a cold dark matter (CDM) particle, the axion, but experiments designed  specifically to detect it via photons have found null results so far. Extensions of the SM, proposed to stabilize the radiative corrections to the Higgs boson mass, commonly have a particle of weak scale mass that is stable because of a discrete symmetry.  Such a weakly interacting massive particle (WIMP), with thermal production in the early universe followed by freeze-out, provides a natural rationale for CDM, with a predicted relic density that is in general accord with its determination from the cosmic microwave background measurements of the Wilkinson Microwave Anisotropy Probe (WMAP) experiment~\cite{Larson:2010gs}. 

There are several means by which DM can be discovered: WIMP pair annihilation in our galactic halo can give positrons, antiprotons, antideuterons, and gamma rays that can be detected in satellite experiments~\cite{Adriani:2008zr,Abdo:2009zk,Aharonian:2009ah,Abdo:2010nc}. Neutrinos from WIMP pair annihilations in the Sun could be observed in large neutrino detectors as events pointing back to the Sun~\cite{Barger:2001ur,Cirelli:2005gh,GonzalezGarcia:2005xw,Barger:2007xf,Barger:2007hj}.  Direct elastic or inelastic scattering of WIMPs can be identified in deep underground detectors via nuclear recoils.  DM may be found at colliders via events with large missing energy carried off by the DM particle.  A concordance of indirect, direct, and collider signals of DM could definitively establish that DM has a particle origin.

A number of direct DM detection experiments are underway and improvements in the upper bounds on the DM scattering cross sections have reached the level sensitivity of interest for DM model tests.  Depending on the choice of the nuclear target, elastic recoils can probe the spin-independent (SI) or spin-dependent (SD) interactions between incoming DM and nucleons.

The Cryogenic Dark Matter Search experiment (CDMS-II) direct detection experiment~\cite{Ahmed:2009zw} completed its five tower run with $612$ kg-days of raw exposure and found two events within the signal region which is consistent with a DM interpretation to $77\%$ C.L.~\cite{Ahmed:2009zw,Bottino:2009km,Cheung:2009wb,He:2009yd,He:2010nt}.  While this observation is not statistically significant, it may be suggestive of a light DM candidate that scatters with low recoil energy.

The DAMA/LIBRA experiment, based on the annual modulation of a DM signal, has found 8.2 sigma  evidence for a low mass DM particle~\cite{Bernabei:2008yi,Bernabei:2010mq,Bottino:2008mf}.  The boundary of the favored signal region is dependent on channeling ~\cite{Savage:2008er,Petriello:2008jj,Feldstein:2009np}, but it has recently been argued that channeling effects are unimportant~\cite{Bozorgnia:2010xy}.  In our study we consider the DAMA/LIBRA boundaries without channeling from Ref.~\cite{Bottino:2009km}.  We also
comment on the DAMA/LIBRA channeled region in the context of the XENON100 exclusion.

The CoGeNT direct DM detection experiment, constructed from $p$-type point contact germanium detectors, benefits from very low electronic noise and high sensitivity to low energy events~\cite{Aalseth:2010vx}.  After rejection of background surface events, an excess of about $100$ low energy, $1.9~\text{keV} < E_{\rm recoil} \lesssim 11$ keV, bulk events was found.  This has been interpreted as the possible signal of  a light DM particle with a mass of $7~\text{GeV} < M_{DM}< 12$ GeV and spin-independent cross section of $\sim 0.7\times 10^{-40}$ cm$^{2}$.  Most of the CoGeNT region is allowed by the CDMS limit.

There have been several model interpretations of the CoGeNT data in terms of a light DM particle~\cite{Andreas:2008xy,Chang:2010yk,Andreas:2010dz,Essig:2010ye,Graham:2010ca}.  Asymmetric DM~\cite{Kaplan:1992db,Hooper:2004dc,Farrar:2005zd,Kitano:2008tk,Kaplan:2009ag,Fitzpatrick:2010em,Cohen:2010kn} connects the DM relic density to the density of baryons, yielding similar DM and baryon masses.  Neutralino DM in the Minimal Supersymmetric Standard Model (MSSM) has tension with $B_s\to \mu^+ \mu^-$, $B^\pm \to \tau \nu$ constraints and sparticle and Higgs boson mass limits~\cite{Bottino:2003cz,Bottino:2003iu,Feldman:2010ke,Kuflik:2010ah}. However, the Next-to-Minimal Supersymmetric Standard Model (NMSSM), a singlet extended MSSM, may provide a viable alternative to vanilla supersymmetry~\cite{Fitzpatrick:2010em}.  The NMSSM model, or any of its supersymmetric cousins, can have Higgs sector phenomenology~\cite{Dermisek:2005gg,Ellwanger:2004gz,Accomando:2006ga,Barger:2006sk,Cerdeno:2008ep,Cerdeno:2009dv} that is somewhat similar to those of the CSM.

Recently the XENON100 collaboration published preliminary DM direct detection exclusion limits~\cite{Aprile:2010um,Collaboration:2010er,Sorensen:2010hq}.  The XENON100 limits with a requirement of $3$ or $4$ photoelectrons (PE) partially overlap with the CoGeNT allowed region.  In our study of the XENON100 constraints we utilize their $4$ PE data.  This exclusion bound may be relaxed somewhat by a different extrapolation of the detection efficiency below $10$ keV nuclear recoil energy~\cite{Collar:2010gg,Collar:2010gd,Sorensen:2010hq}, in which case the XENON100 limits could even be consistent with the full CoGeNT allowed region.  

In this paper, we consider an explanation of either the CoGeNT DM signal, the DM signal from the DAMA/NaI and DAMA/LIBRA data, or the XENON100 DM allowed region,  in the context of the complex scalar singlet extended standard model (CSM)~\cite{Barger:2008jx}.   The complex singlet provides two additional degrees of freedom, one which acts as a scalar singlet that mixes with the SM-Higgs boson and the other a DM candidate whose stability is ensured by CP conservation of the scalar potential.  The real scalar singlet model ~\cite{McDonald:1993ex,Burgess:2000yq,BahatTreidel:2006kx,OConnell:2006wi,Barger:2007im,Asano:2010yi,Arina:2010an,Bandyopadhyay:2010cc} can provide a dark matter candidate but then has only a SM-Higgs boson.

In Section~\ref{sect:model}, we present an overview of the complex singlet extended SM and its features.   The experimental constraints on the parameters of the model are described in Section~\ref{sect:constraints}.  Sections~\ref{sect:sipmeasure} and~\ref{sect:higgs} give results where we show the correlated signatures in the Higgs sector that are a consequence of the light DM particle with a spin-independent scattering cross section consistent with CoGeNT, DAMA/LIBRA and/or XENON100.  In Section~\ref{sect:conclusions}, we provide conclusions and an outlook.

%%%%%%%%%%%%%%%%%%%%%%%%%%%%%%%%%%
\section{The CSM Model}
\label{sect:model}
%%%%%%%%%%%%%%%%%%%%%%%%%%%%%%%%%%

In the CSM model, the singlet fields talk to the SM only via the Higgs boson.  Thus, the DM sector of the model is a concrete realization of the Higgs portal concept~\cite{Patt:2006fw,Arina:2010an}. The scalar potential of the CSM, including only renormalizable terms is
\begin{eqnarray}
V_\mathrm{CSM}&=&\frac{m^{2}}{2}H^{\dagger}H+\frac{\lambda}{4}(H^{\dagger}H)^{2}+\frac{\delta_{2}}{2}H^{\dagger}H|\cs|^{2}+\frac{b_{2}}{2}|\cs|^{2}+\frac{d_{2}}{4}|\cs|^{4} \nonumber \\
&+&\left(\frac{|b_{1}|}{4}e^{i\phi_{b_1}}  \cs^{2}+|a_1|\, e^{i\phi_{a_1}} \cs + c.c.\right),
\label{eq:potential}
\end{eqnarray}

\noindent where $H$ is the SM-Higgs and $\cs = (S+iA)/\sqrt{2}$ is the complex singlet.  Starting with a global $U(1)$ symmetric potential we include a $U(1)$ breaking term $b_{1}$~\cite{Barger:2008jx}.  In the case that the real component of the singlet obtains a vacuum expectation value (vev), a nonzero  $a_{1}$ is needed to avoid domain walls from the accidental $\cs\to-\cs$ symmetry.  In fact, we find that nonzero $a_{1}$ is essential to describe the experimental observations, discussed in the next section. The phases $\phi_{b_{1}}$ and $\phi_{a_{1}}$ need to be either $0$ or $\pi$ in order to avoid mixing between the real and complex components of $\cs$ and thus provide a stable DM candidate, $A$, as we desire.  Defining $a_{1} $ and $b_{1}$ to be positive definite, we find that the phases must be $\phi_{b_{1}}=\pi$ and $\phi_{a_{1}}=0$ for viable phenomenology~\cite{Barger:2008jx}. 

If the real component of the complex singlet obtains a vev, $v_{S}$, the $\delta_{2}$ term in the potential initiates mixing between $H$ and $S$ resulting in two Higgs particles with masses
\begin{equation}
M_{H_{1,2}}^{2}=\frac{1}{4}\left(\lambda v^{2}+d_{2}v_{S}^{2}-\frac{2\sqrt{2}a_{1}}{v_{S}}\mp\sqrt{\left(\lambda v^{2}-d_{2}v_{S}^{2}+\frac{2\sqrt{2}a_{1}}{v_{S}}\right)^{2}+4\delta_{2}^{2}v^{2}v_{S}^{2}}\right).
\label{eq:mass}
\end{equation}
The resulting coupling strengths of these Higgs eigenstates to the SM fermions and weak bosons are multiplied by the factors
\begin{equation}
g_{H_{i}}=\left\{
\begin{array}{ccc}
\cos\phi\quad \mathrm{for }~H_{1}\\
-\sin\phi\quad \mathrm{for }~H_{2}
\end{array}
\right.,
\end{equation}
where the mixing angle is given by
\begin{equation}
\tan2\phi=\frac{2\delta_{2}vv_{S}}{\lambda v^{2}-d_{2}v_{S}^{2}+\frac{2\sqrt{2}a_{1}}{v_{S}}}.
\label{eq:mixing}
\end{equation}
The remaining complex term in $V$ leads to a scalar field $A$ that is stable and is  thus the DM candidate.  The mass of the DM particle is determined by the parameters $b_{1}$ and $a_{1}$
\begin{equation}
M_{A}^{2} = b_{1}-\frac{\sqrt{2}a_{1}}{v_{S}}.
\end{equation}

%%%%%%%%%%%%%%%%%%%%%%%%%%%%%%%%%%
\section{Constraints}
\label{sect:constraints}
%%%%%%%%%%%%%%%%%%%%%%%%%%%%%%%%%%

The CSM potential [Eq.~\ref{eq:potential}] has six free parameters: the SM-Higgs quartic coupling $\lambda$, the complex singlet quartic coupling $d_{2}$, the quartic interaction between the SM-Higgs and complex singlet $\delta_{2}$, the vev of the real component of the complex singlet $v_{S}$, and the DM mass parameters $b_{1}$ and $a_{1}$.  The $m^{2}$ parameter in Eq.~\ref{eq:potential} is determined in terms of the other parameters by the minimization conditions of the potential.

We take the quartic couplings to be $\mathcal{O}(1)$ for perturbativity and we allow the singlet vev to be as large as $1$ TeV.  We scan uniformly over the following ranges:
\begin{eqnarray}
7~\text{GeV}~<&M_{A}&<~1000~\text{GeV},\\
M_{A}~<&\sqrt{b_{1}}&<~10M_{A},\\
0~<&\lambda&<~2,\\
-2~<&\delta_{2}&<~2,\\
0~<&d_{2}&<~4,\\
10~\text{GeV}~<&v_{S}&<~1000~\text{GeV},
\end{eqnarray}

Note that the range of $\sqrt{2}a_{1}/v_{S}$ is determined by the ranges of $M_{A}$ and $b_{1}$.  Additionally, each parameter set is required to have
\begin{equation}
M_{H_{i}}^2>0,
\end{equation}
as having a stable global vacuum demands~\cite{Barger:2008jx}.

\noindent{\bf Relic Density --}
We require the DM candidate to fully saturate the relic abundance of thermal DM within a $2\sigma$ range of the value measured by the WMAP experiment~\cite{Larson:2010gs}:
\begin{equation}
0.098 < \Omega_{A}h^{2} < 0.122
\end{equation}
This is a very restrictive condition but we find that it can be satisfied in conjunction with the DM experimental constraints.  There are also multiple solutions with an under saturated DM density, which could be relevant if there is another DM contributor (such as the axion), or over saturated, which is possible in nonstandard cosmologies, such as a very low reheating temperature or extra entropy injection caused by late decays~\cite{Gelmini:2006pw,Salati:2002md,Profumo:2003hq,Rosati:2003yw,Chung:2007vz}.

\begin{figure}[htbp]
\begin{center}
\includegraphics*[angle=0,width=0.19\textwidth]{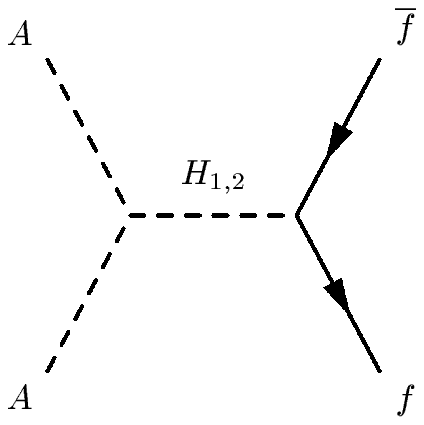}~~~
\includegraphics*[angle=0,width=0.16\textwidth]{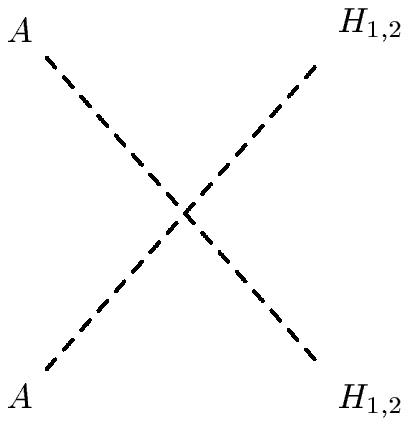}~~~
\includegraphics*[angle=0,width=0.16\textwidth]{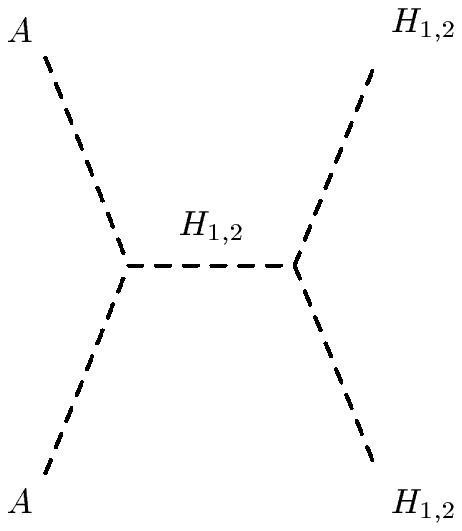}~~~
\includegraphics*[angle=0,width=0.16\textwidth]{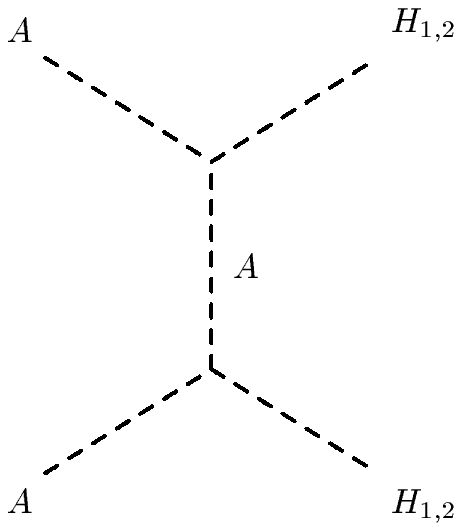}~~~
\includegraphics*[angle=0,width=0.19\textwidth]{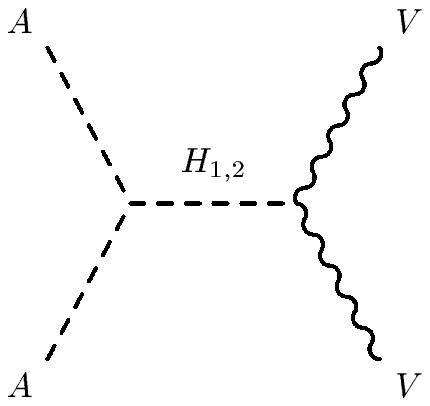}
\caption{Feynman diagrams that contribute to the annihilation cross setion of $A$.  All processes are mediated via the two Higgs eigenstates $H_{1,2}$. }
\label{fig:annFD}
\end{center}
\end{figure}

The Feynman diagrams associated with DM annihilation are given in Fig.~\ref{fig:annFD}.  As noted previously, the only communication between the DM and SM is through the Higgs mass eigenstates.  In the putative DM mass region inferred by CoGeNT, is $AA\to b\bar{b}$, since the $b$-quark is the heaviest kinematically accessible fermion to which the virtual Higgs couples. For a DM mass that is below the threshold of the $b\bar{b}$ channel, the DM annihilates to $c\bar{c}$ and $\tau^{+}\tau^{-}$ final states.   Occasionally, $H_1$ is light enough that the $AA\to H_{1} H_{1}$ channel may contribute to the relic abundance.  To calculate the relic density, we utilize the micrOMEGAs software package~\cite{Belanger:2008sj,Belanger:2010gh}. 

\noindent{\bf LEP data --}
The OPAL collaboration~\cite{Abbiendi:2002qp} has placed constraints on the production of scalar particles in the $Z$-Higgs-strahlung process, independent of their decay modes, expressed in terms of bounds on a quantity $k_{i}$ defined as
\begin{equation}
k_{i}=\frac{\sigma(e^{+}e^{-}\to ZH_{i})}{\sigma(e^{+}e^{-}\to Zh_{SM})}.
\end{equation}

It is possible to have Higgs eigenstates that are lighter than the LEP limit of $114$ GeV on the SM-Higgs mass, since the mixing between the SM-Higgs and scalar singlet field reduces the couplings.  We impose the LEP bounds, that exclude a SM-Higgs boson of mass $12-114$ GeV~\cite{Sopczak:2005mc}, to both $ZZH_{i}$ couplings.  In addition, the combined LEP data place upper bounds on the quantities $\xi_{i}^{2}$ defined as
\begin{equation}
\xi_{i}^{2}=k_{i}\times \text{BF}(H_{i}\to SM).
\end{equation}
Because of the mixing of the SM-Higgs with the real component of the complex singlet and possible new Higgs decay modes to DM particles, the Higgs masses can be lower than the LEP bound on the SM-Higgs mass, depending on the mixing:  For an illustration of this, see Fig. 6c.  In addition to the limit on the standard search modes, we apply the DELPHI~\cite{Abdallah:2003ry} and combined LEP~\cite{LEP:2001xz} limit on the invisible decay of a Higgs state, where a SM-Higgs boson must have mass $m_H > 114$ GeV for $\xi_i^2=1$.

\noindent{\bf Higgs Cascade Decays --}
With the two Higgs mass eigenstates in the CSM, there is the possibility of cascade decays, i.e. $H_{2}\to H_{1}H_{1}\to f\bar f f^\prime \bar f^\prime$.  Such decay chains have been considered in the context of $b\bar{b}+b\bar{b}$ and $b\bar{b}+\tau^{+}\tau^{-}$~\cite{Dermisek:2005gg,Barger:2006sk,Dermisek:2006wr,Carena:2007jk,Cheung:2007sva,Chang:2008cw}.  These decay channels are also constrained by the decay-independent bounds obtained by OPAL from $Z$-boson recoils~\cite{Abbiendi:2002qp}.  

The Higgs cascade decay cross sections, relative to the total SM $Z$-Higgs-strahlung cross section, are given by
\begin{eqnarray}
C_{b\bar b + b\bar b} &=&k_{i}\times\text{BF}(H_{2}\to H_{1}H_{1})\times\text{BF}(H_{1}\to b \bar b)^{2}\\
C_{\tau^+\tau^- + \tau^+\tau^-} &=&k_{i}\times\text{BF}(H_{2}\to H_{1}H_{1})\times\text{BF}(H_{1}\to\tau^{+}\tau^{-})^{2}\\
C_{b\bar b + \tau^+\tau^-} &=&2k_{i}\times\text{BF}(H_{2}\to H_{1}H_{1})\times\text{BF}(H_{1}\to\tau^{+}\tau^{-})\times\text{BF}(H_{1}\to b\bar b).
\end{eqnarray}

The LEP collaboration has placed upper limits on these quantities.  The $b\bar b+b\bar b$ channel is limited to $C_{b\bar b + b\bar b}<0.2$ for masses $M_{H_2}\lesssim 85$ GeV, while for $M_{H_2}\approx 105\text{ GeV}$, the limit relaxes to $C_{b\bar b + b\bar b}<0.4$~\footnote{In our scans, we typically find $M_{H_2}\gtrsim110$ GeV after all other constraints are applied.}.  The $b\bar b +\tau^+\tau^-$ modes are limited to be $C_{b\bar b + \tau^+\tau^-}<0.4$ for masses $M_{H_2}\lesssim 85$ GeV while $C_{b\bar b + \tau^+\tau^-}$ can be as large as unity for masses $M_{H_2}\gtrsim 105$ GeV.  In addition, we include the limits on the $\tau^{+}\tau^{-}+\tau^{+}\tau^{-}$ decays from the ALEPH collaboration: $C_{\tau^+\tau^-+\tau^+\tau^-}>1$  for values of $M_{H_{2}}>107$ GeV and $4$ GeV $<M_{H_1}<10$ GeV at the $95\%$ C.L.~\cite{Aleph:2010aw}.  In our scans, we rarely find values exceeding $C_{b\bar b + \tau^+\tau^-}=0.12$ and $C_{\tau^+\tau^- + \tau^+\tau^-}=0.1$.  Therefore, the $b\bar b+\tau^+\tau^-$ and $\tau^+\tau^- + \tau^+\tau^-$ constraints are not very effective in excluding parameter regions.

\noindent {\bf Electroweak Precision Observables (EWPO) --}
The constraints from EWPO require that the mass of a SM-like Higgs must be bounded by $M_{h}\lesssim180$ GeV.  In the CSM this applies to the mass of the heavier Higgs eigenstate~\cite{Amsler20081}.

\begin{figure}[htbp]
\begin{center}
\includegraphics*[angle=0,width=0.18\textwidth]{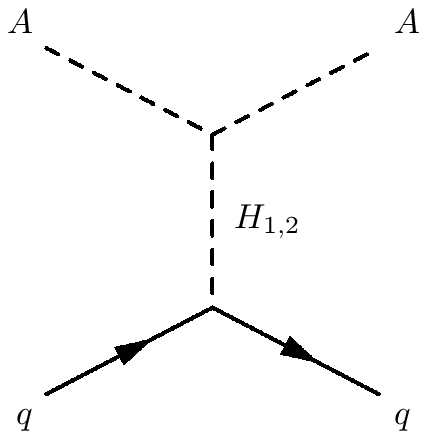}~~~
\includegraphics*[angle=0,width=0.18\textwidth]{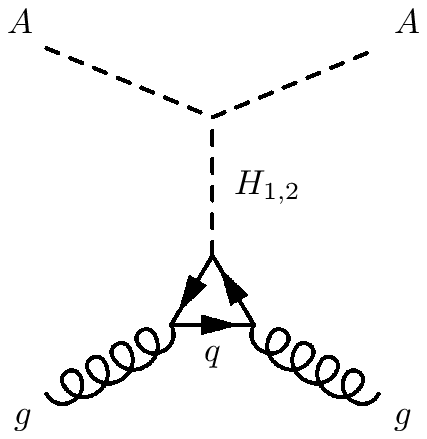}
\caption{Feynman diagrams associated with the spin-independent direct detection of $A$.  As in annihilations, the Higgs bosons are the mediating fields between the DM and the nuclear target. }
\label{fig:sidd}
\end{center}
\end{figure}

\noindent{\bf $\sigma_{SI}$ Measurements  --} The Feynman diagrams associated with the spin-independent scattering cross section are shown in Fig.~\ref{fig:sidd}.  For the CSM, the spin-independent scattering cross section of DM on a proton target is
\begin{equation}
\sigma_{SI}=\frac{m_{p}^{4}}{2\pi v^{2}(m_{p}+M_{A})^{2}}\left(\frac{g_{AAH_{1}}g_{H_{1}}}{M_{H_{1}}^{2}}+\frac{g_{AAH_{2}}g_{H_{2}}}{M_{H_{2}}^{2}}\right)^{2}\left(f_{pu}+f_{pd}+f_{ps}+\frac{2}{27}(3f_{G})\right)^{2},
\label{eq:sip}
\end{equation}
\noindent where $m_{p}$ is the proton mass and $v=246$ GeV is the SM-Higgs vev.  The couplings between the DM particles and the Higgs eigenstates can be written as
\begin{eqnarray}
g_{AAH_{1}}&=&\left(\delta_{2}v\cos\phi+d_{2}v_{S}\sin\phi\right)/2,\\
\label{eq:gh1aa}
g_{AAH_{2}}&=&\left(d_{2}v_{S}\cos\phi-\delta_{2}v\sin\phi\right)/2.
\label{eq:gh2aa}
\end{eqnarray}
\noindent The strengths of the hadronic matrix elements, $f$, have the central values~\cite{Ellis:2000ds}
\begin{equation}
f_{pu} = 0.02,\qquad f_{pd} = 0.026,\qquad f_{ps} = 0.118,\qquad f_{G}=0.836.
\end{equation}
The cross sections calculated using Eq.~\ref{eq:sip} agree with those calculated by micrOMEGAs~\cite{Belanger:2008sj,Belanger:2010gh}.  We do not consider uncertainties on the hadronic matrix elements in our analyses, as is the standard practice.  In general, the variation of these values can shift the overall scattering rate by roughly $\pm30\%$~\cite{Barger:2008qd}.

\noindent The expression for the SI cross section can be simplified to
\begin{equation}
\sigma_{SI}=\frac{m_{p}^{4}}{2\pi (m_{p}+M_{A})^{2}}\left( { \delta_2 (b_1-M_A^2) \over 2 M_{H_{1}}^{2} M_{H_{2}}^{2} }\right)^{2} \left(f_{pu}+f_{pd}+f_{ps}+\frac{2}{27}(3 f_{G})\right)^{2}.
\end{equation}  

\noindent We note that $\sigma_{SI}$ vanishes in the limits $\delta_{2}\to0$ or $b_{1}\to M_A^2$.  In terms of representative masses and parameters, the cross section is
\begin{equation}
\sigma_{SI}\approx1.5\times10^{-5}\text{pb}\left(\frac{100~\text{GeV}}{M_{H_{1}}}\right)^4\left(\frac{100~\text{GeV}}{M_{H_{2}}}\right)^{4}\left(\frac{\delta_{2}(b_1-M_A^2)}{10^{3}~\text{GeV}^{2}}\right)^{2}.
\label{eq:sipapprox}
\end{equation}
In our subsequent  CSM study, we separately consider the CoGeNT signal region , the DAMA/LIBRA signal region, and the XENON100 exclusion region.  

%%%%%%%%%%%%%%%%%%%%%%%%%%%%%%%%%
\section{Consistency with $\sigma_{SI}$ measurements}
\label{sect:sipmeasure}
%%%%%%%%%%%%%%%%%%%%%%%%%%%%%%%%%

We can correlate the $\sigma_{SI}$ signal with the scalar mass patterns in the CSM.  Mass ranges that give rise to characteristic phenomenological features are
\begin{enumerate}
\item $M_{H_{1}} < M_A$.--the annihilation $AA\to H_{1}H_{1}$ is efficient enough to saturate the relic density.  These points are denoted by black crosses in the figures.
\item $M_A < M_{H_{1}} < 2 M_A $.--here $2M_A$ is just above the $H_{1}$ resonance and below the $H_1 H_1$ threshold, for nonrelativistic $A$ (closed red circles).  With additional thermal energy of the $A$ in the early universe, the $AA\to H_1 H_1$ mode may be open for $M_A\lesssim M_{H_1}$.
\item $ M_{H_{1}} > 2 M_A $.--the $A$ mass is just below the $H_{1}$ resonance, hereafter denoted by open blue boxes.
\end{enumerate}

\begin{figure}[b]
\begin{center}
\includegraphics*[angle=0,width=0.6\textwidth]{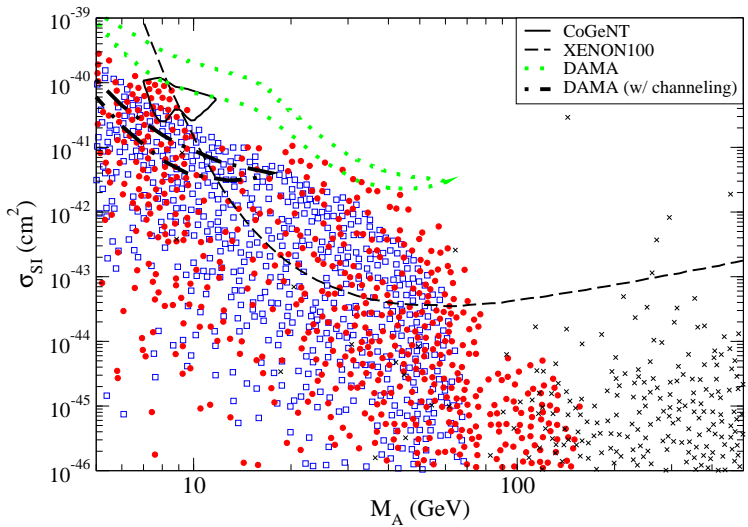}
\caption{Range of $M_A$ and $\sigma_{SI}$ values.  The 90\% C.L. boundary of the CoGeNT signal region is denoted by the solid (black) contour.  The signal region of the DAMA/LIBRA data, with no channeling effects, is enclosed by a dotted (green) contour.  The XENON100 exclusion limit is given by the short-dashed (black) boundary. The DAMA/LIBRA region with channeling is shown by the bold dash-dotted (black) contour. All contours are shown at the $90\%$ C.L.  The measured relic density can be saturated below the $H_{1}$ resonance (open blue boxes), above the $H_{1}$ resonance (filled red circles), with the $AA\to H_{1}H_{1}$ channel open during freeze-out (black crosses). }
\label{fig:sipmagen}
\end{center}
\end{figure}
 
In Fig.~\ref{fig:sipmagen}, we present the ranges of $M_A$ and $\sigma_{SI}$ after all other constraints are applied.   The CoGeNT signal region at $90\%$ C.L. is denoted by the solid (black) bounding curve.  The DAMA/LIBRA signal region with channeling effects is within the bold dash-dotted (black) boundary while the signal region without channeling effects is enclosed within the dotted (green) boundary.  The $90\%$ exclusion region from XENON100 ($4$ PE) is denoted by the dashed (black) curve.  The CSM points correspond to $M_{H_{1}} < M_A$: crosses (black), $M_A < M_{H_{1}} < 2 M_A $ filled circles (red), and $ M_{H_{1}} > 2 M_A $ $ M_{H_{1}} > 2 M_A $.

We see that the CSM points populate the low $M_A$ part of the CoGeNT region.  

Only a few points are consistent with the DAMA/LIBRA region without channeling.  One group of points lies along the lower DAMA/LIBRA boundary at $M_A \sim 30$ GeV, which is consistent with XENON100.  The other lies at $M_A \sim 7$ GeV, for which the phenomenology will be similar to that of the CoGeNT region; therefore, we do not explicitly list these points but note the conclusions are similar to those obtained from analysis of the CoGeNT region, except that the SI cross section is slightly higher.  More model points in these two DAMA/LIBRA consistent regions could be realized if the uncertainties discussed earlier in the calculated cross sections were taken into consideration.  The XENON100 limit convincingly excludes the DAMA/LIBRA consistent unchanneled region at $M_A \sim 30$ GeV, but  it allows a wide mass range from $10$ GeV to the TeV scale.  In our more detailed considerations below we focus our discussion on the CoGeNT region at $M_A\sim 10$ GeV and the region that is not excluded by XENON100.  A portion of the channeled DAMA/LIBRA allowed region is consistent with the XENON100 exclusion.  We note that the channeling calculations are based on complex modeling, so it is possible that the real situation could lie between the unchanneled and channeled regions shown in Fig.~\ref{fig:sipmagen}.  Therefore, small channeling effects would improve the agreement of the DAMA/LIBRA data with the model.

\begin{figure}[htpb]
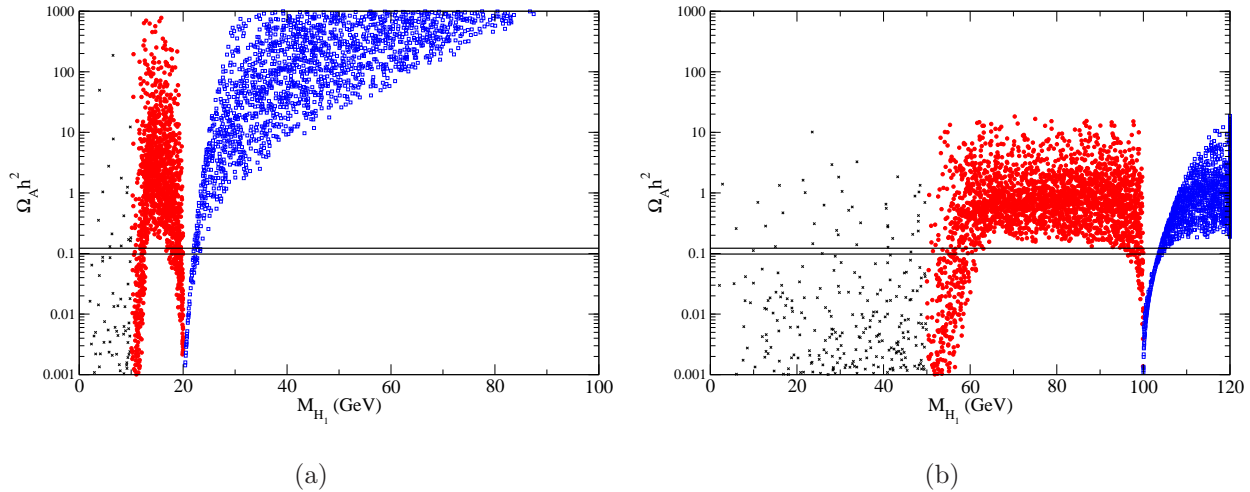

\begin{center}
\subfigure[]{\includegraphics*[angle=0,width=0.49\textwidth]{figs/oh2vsmh1-cogent.eps}}~~
\subfigure[]{\includegraphics*[angle=0,width=0.49\textwidth]{figs/oh2vsmh1-xenon.eps}}\\
\caption{Relic abundance of $A$ versus the $H_{1}$ mass with $M_{H_{2}} = 120$ GeV and a $H_{1}$ SM-Higgs content of $10^{-4}$ in (a) the CoGeNT region with $M_{A}=10$ GeV and (b) XENON100 exclusion region with $M_{A}=50$ GeV.  The measured relic density can be saturated below the $H_{1}$ resonance (blue open boxes), above the $H_{1}$ resonance (red filled circles), with the $AA\to H_{1}H_{1}$ channel open during freeze-out (black crosses). This color notation is respected in subsequent figures unless otherwise noted.}
\label{fig:hpole}
\end{center}
\end{figure}

In Fig.~\ref{fig:hpole}, for a DM mass of $M_A=10$ GeV, representative of CoGeNT and DAMA/LIBRA, and a $H_{1}$ SM-Higgs content of $10^{-4}$ we identify the light Higgs masses for which DM annihilation reproduces the observed DM relic density.  There are a variety of ways in which the relic density may be satisfied and fall into the three categories outlined above.  When $M_{A}<M_{H_{1}}<2M_{A}$, there are two possible ways to saturate the relic density: above the $AA\to H_{1}$ resonance, shown as the right branch of case (ii), or above the $AA\to H_{1}H_{1}$ threshold.  When DM annihilates above the $H_1$ pair threshold, the thermal energy in the early universe is large enough to open that channel which would otherwise be closed in the $v_{ann}\to 0$ limit.  In Fig.~\ref{fig:hpole}b, we show the results for $M_A=50$ GeV, the DM mass for which XENON100 is most sensitive.  Generally, these regions may be shifted down in cross section as the values of the SM-Higgs content of the $H_1$ state increases.

\begin{figure}[htpb]
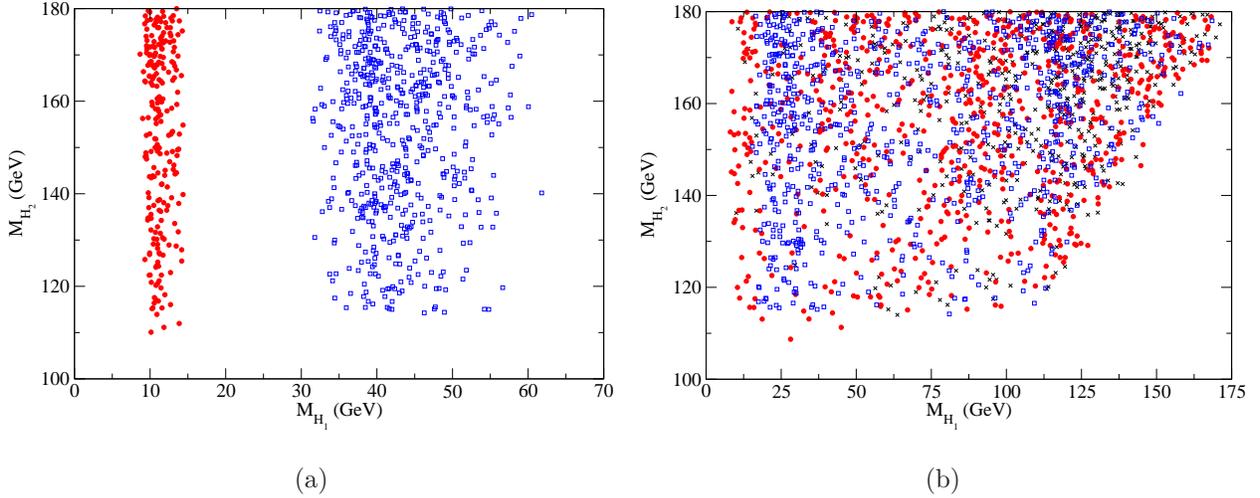

\begin{center}
\subfigure[]{\includegraphics*[angle=0,width=0.49\textwidth]{figs/mh2vsmh1-cogent.eps}}~~~
\subfigure[]{\includegraphics*[angle=0,width=0.49\textwidth]{figs/mh2vsmh1-xenon.eps}}
\caption{Mass of the heavy Higgs versus that of the light Higgs for (a) CoGeNT $90\%$ C.L. signal region and (b) XENON100 allowed region.  The $AA$ annihilation takes place above (below) the $H_1$ resonance for the blue boxes (red circles).}
\label{fig:mh2mh1}
\end{center}
\end{figure}

With all of the aforementioned constraints applied to our parameter scan, we find that in the CoGeNT case there are no points with $M_{H_{1}}<M_{A}$ that survive all of the constraints.  The mass ranges of the light Higgs are $9-15$ GeV and $30-65$ GeV, depending on which side of the $H_1$ resonance the $A$ state lies.   The mass of the heavy Higgs is $110-180$ GeV: see Fig.~\ref{fig:mh2mh1}. The upper bound on the light Higgs mass corresponds to the upper bound on the relic density.  If $M_{H_{1}}$ is significantly greater than twice the DM mass, the $AA\to f\bar{f}$ annihilation cross section is suppressed by $m_{f}^{2}/M_{H_{1}}^{4}$, producing an over abundance of relic DM.

\begin{figure}[htpb]
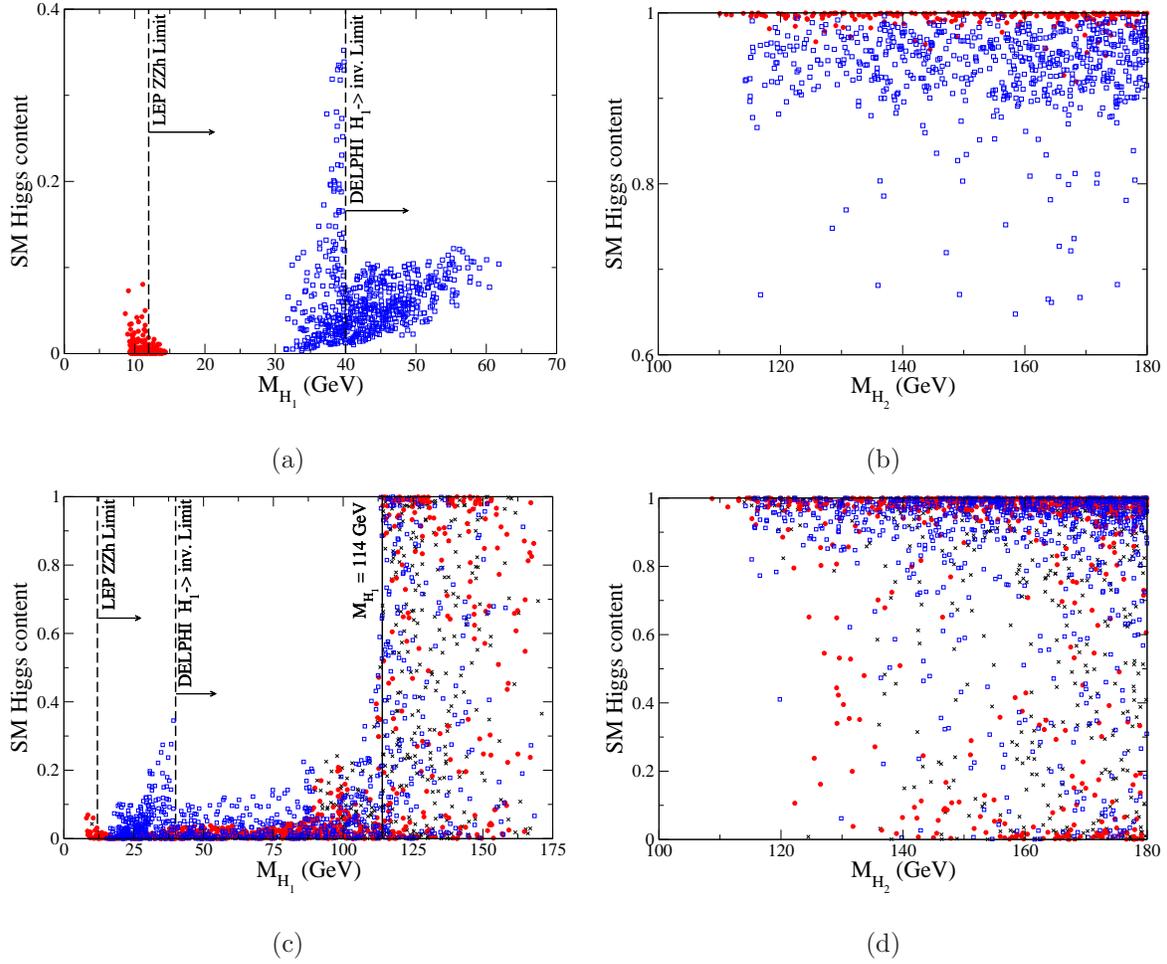

\begin{center}
\subfigure[]{\includegraphics*[angle=0,width=0.45\textwidth]{figs/costhetavsmh1-cogent.eps}}~~~
\subfigure[]{\includegraphics*[angle=0,width=0.45\textwidth]{figs/sinthetavsmh2-cogent.eps}}
\subfigure[]{\includegraphics*[angle=0,width=0.45\textwidth]{figs/costhetavsmh1-xenon.eps}}~~~
\subfigure[]{\includegraphics*[angle=0,width=0.45\textwidth]{figs/sinthetavsmh2-xenon.eps}}
\caption{The SM-Higgs probability content of the Higgs eigenstates consistent with the CoGeNT data (a-b) and XENON100 data (c-d).  Lighter Higgs bosons are dominantly singlet, largely due to the LEP $ZZH_{i}$ mixing constraints.  For larger $H_{1}$ masses the $AA$ decay is kinematically allowed and this diminishes the branching fraction of $H_{1}$ to SM channels}
\label{fig:costhetamh1}
\end{center}
\end{figure}

Because of the mixing of the SM-Higgs field and the singlet field, the $H_2$ mass may be below the SM-Higgs limit given by LEP.  Therefore, any Higgs states with masses much below the 114 GeV LEP limit are dominantly singlet.  This can be easily seen in Fig.~\ref{fig:costhetamh1}.  The CoGeNT consistent cases (top panels) give a light singlet due to the relic density constraint, forcing the heavier state to be SM-like.  In the XENON100 cases (bottom panels), there is no such restriction, allowing both mass states to have a relatively free SM-Higgs component.  The abrupt shifts in the limit of the SM-Higgs content are caused by the endpoint of the constraining data as indicated.

\begin{figure}[htpb]
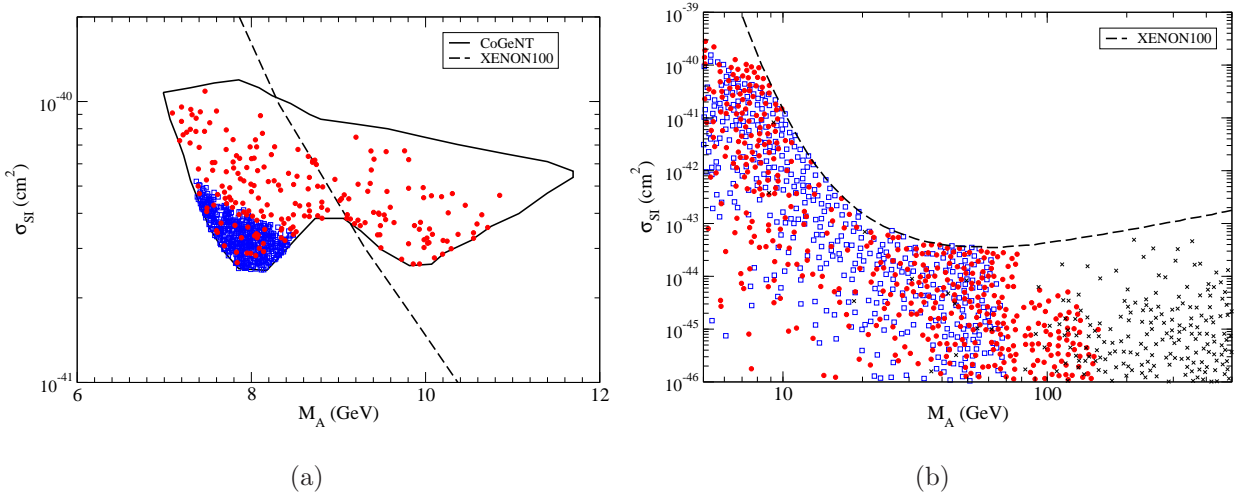

\begin{center}
\subfigure[]{\includegraphics*[angle=0,width=0.49\textwidth]{figs/sipvsma-cogent.eps}}~
\subfigure[]{\includegraphics*[angle=0,width=0.49\textwidth]{figs/sipvsma-xenon.eps}}
\caption{SI cross section versus the DM mass in the CSM model within (a) the $90\%$ C.L. boundary of the CoGeNT region with the XENON100 $90\%$ C.L. exclusion overlain and (b) the XENON100 allowed region, see the text for details.  All model points satisfy the measured WMAP relic abundance.  }
\label{fig:crossvsmdm}
\end{center}
\end{figure}

Finally, using Eq.~\ref{eq:sip}, we obtain the results in Fig.~\ref{fig:crossvsmdm} after all the constraints are imposed.  We find that points with $M_{H_{1}}>2M_{A}$ prefer smaller $M_{A}$ in order to satisfy the CoGeNT bounds and the observed WMAP relic density while the points where $M_{A}<M_{H_{1}}<2M_{A}$ populate a larger area of the CoGeNT boundary.  Generally, with the light Higgs mass relatively small ($9-15$ GeV), there is more freedom in the range of the spin-independent scattering cross section: see Eq.~\ref{eq:sipapprox} and Fig.~\ref{fig:mh2mh1}.  The XENON100 $90\%$ C.L. consistent regions possess a continuum of $H_1$ pole solutions from the lower limit to $\sim90-110$ GeV due to the maximum value of the Higgs boson masses by EWPO constraints.  Above this $H_1$ pole wall, the $AA\to H_1 H_1$ channels are open, generally providing a scattering cross section that is well below the XENON100 exclusion limit.

For comparison, we superimpose the $90\%$ C.L. XENON100 exclusion limit with $4$ PE on top of the $90\%$ C.L. CoGeNT region.  The lower left region is in best agreement with the XENON100 limit.  It is also in this region that we find the best agreement with the CSM.

%%%%%%%%%%%%%%%%%%%%%%%%%%%%%%%%%%
\section{Higgs signatures}
\label{sect:higgs}
%%%%%%%%%%%%%%%%%%%%%%%%%%%%%%%%%%

The imposition of the DM relic density constraint and the CoGeNT constraint on the DM scattering cross section allow firm predictions to be made about the Higgs sector of this model.  The mass configurations of the three scalar states largely determine the dominant decay modes of the SM-like $H_{2}$ state. 
\begin{figure}[htpb]
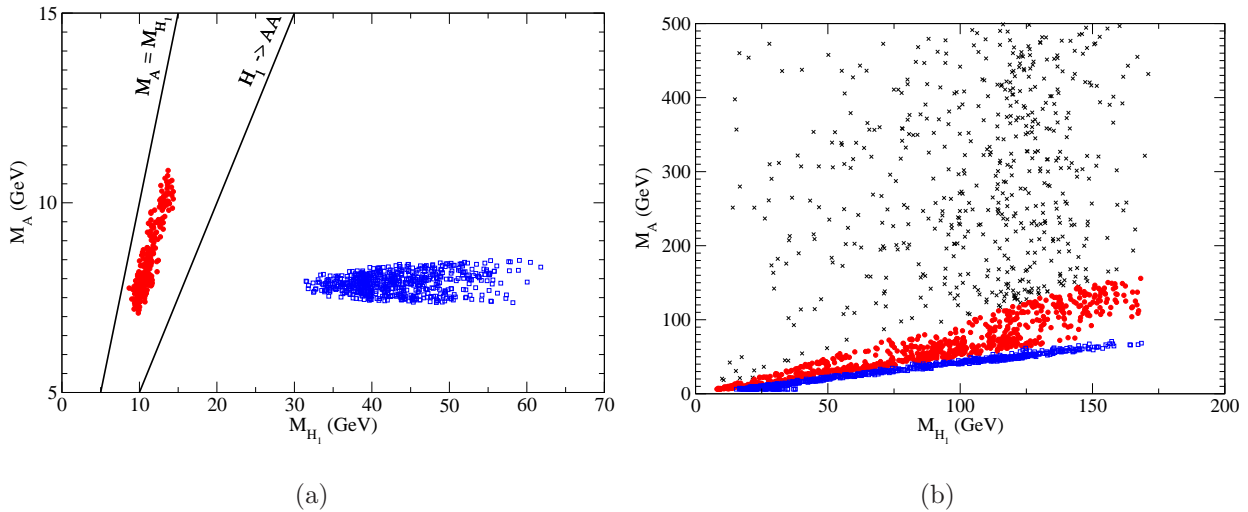

\begin{center}
\subfigure[]{\includegraphics*[angle=0,width=0.49\textwidth]{figs/mh1vsma-cogent.eps}}~
\subfigure[]{\includegraphics*[angle=0,width=0.49\textwidth]{figs/mh1vsma-xenon.eps}}
\caption{DM mass versus the light Higgs mass for (a)  the $90\%$ C.L. boundary of the CoGeNT allowed region and (b) the XENON100 $90\%$ C.L. consistent region.  The $H_{1}\to AA$ decay mode is open and dominant for $M_A < \half M_{H_1}$ (blue open boxes).}
\label{fig:mh1vsma}
\end{center}
\end{figure}

In Fig.~\ref{fig:mh1vsma}, we show the $A$ mass versus the $H_{1}$ mass after all constraints are applied.  Three regions which control the dominant annihilation process for $AA\to \text{SM}+\text{SM}$ that sets the relic density become apparent.  Below (boxes) the $AA\to H_{1}\to \text{SM}+\text{SM}$ resonance, the relic density constraint can be satisfied.  Additionally, the $AA\to H_{1}H_{1}$ mode is open when $M_A>M_{H_{1}}$ (crosses).  Below the solid line, the $H_{1}\to AA$ mode is open and can be a source of $\met$ at hadron colliders.  For all of the CoGeNT and many of the XENON100 cases, $H_{2}\to H_{1}H_{1}$ decays are kinematically accessible, possibly yielding a dominant decay mode of the $H_2$ Higgs boson.

Overall, there are three main decay processes connected with the Higgs mass spectra as follows~\footnote{In some cases a $H_{2}\to 3H_{1}$ decay is kinematically allowed.  Because of the small coupling and phase space suppression this mode will not be a dominant decay mode}:
\begin{enumerate}
\item Large decay rates of $H_{2}\to$ SM-modes.  Predominantly, these modes are $W^+W^-$ and $ZZ$ due to the rapid growth of these partial widths with the Higgs mass, $\Gamma\sim M_{H_{2}}^3/M_W^2$.  When the $W^+W^-$ mode is fully open, $M_{H_{2}}> 2 M_W$, the SM width increases rapidly with $M_{H_{2}}$ and it is comparable to or dominates over the nonstandard decays.  The effective production rates of the typical SM modes through the SM-like Higgs boson, $H_{2}$, are shown in Fig.~\ref{fig:BFh2SMcogent} for the CoGeNT DM region and Fig.~\ref{fig:BFh2SMxenon} for the XENON100 exclusion region.  For both regions, we see a reduction from the SM branching fraction.  For the CoGeNT region, this is primarily due to the other decay modes listed below that compete with the SM decay modes, while for the XENON100 region, the significant singlet content of the $H_2$ state can provide an additional level of suppression.  In addition, the XENON100 region may have a SM-like Higgs that cannot decay to a singlet state that can have a larger mass than in the CoGeNT case, allowing the branching fraction to be closer to the SM expectation.

\begin{figure}[htpb]
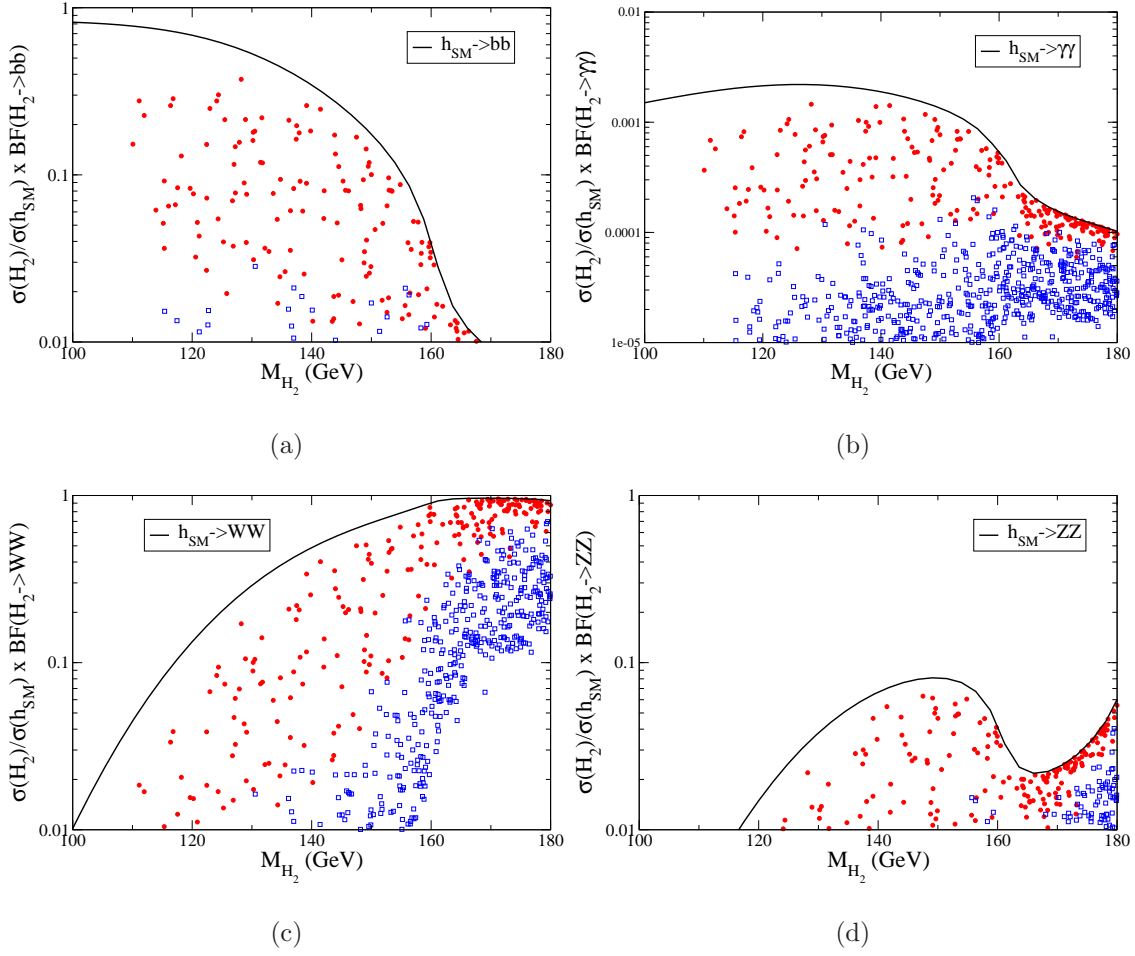

\begin{center}
\subfigure[]{\includegraphics*[angle=0,width=0.45\textwidth]{figs/bfh2bb-cogent.eps}}
\subfigure[]{\includegraphics*[angle=0,width=0.45\textwidth]{figs/bfh2gam-cogent.eps}}\\
\subfigure[]{\includegraphics*[angle=0,width=0.45\textwidth]{figs/bfh2ww-cogent.eps}}
\subfigure[]{\includegraphics*[angle=0,width=0.45\textwidth]{figs/bfh2zz-cogent.eps}}
\caption{Effective production rates of a (a) $b\bar b$, (b) $\gamma \gamma$, (c) $W^+W^-$, and (d) $ZZ$ through the heavy Higgs boson versus its mass for the $90\%$ C.L. CoGeNT allowed region.  Suppression with respect to the SM expectation can largely be obtained via competing decay modes $H_2\to H_1 H_1$ and $H_2\to AA$. Recent Tevatron exclusion limits on $H\to W^{+}W^{-}$ are beginning to impact a large $WW$ branching fraction in the Higgs mass range $160-200$ GeV~\cite{Aaltonen:2010sv}.}
\label{fig:BFh2SMcogent}
\end{center}
\end{figure}

\begin{figure}[htpb]
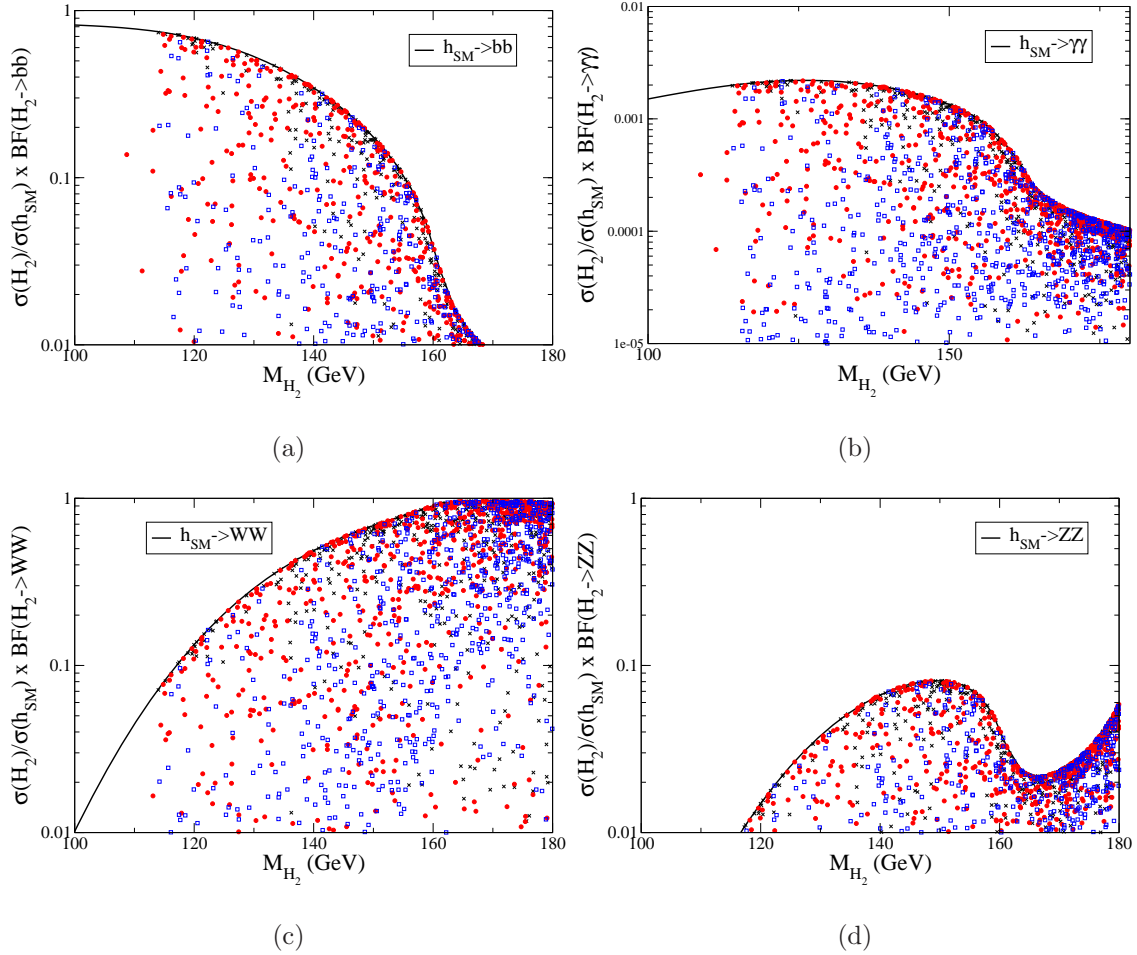

\begin{center}
\subfigure[]{\includegraphics*[angle=0,width=0.45\textwidth]{figs/bfh2bb-xenon.eps}}
\subfigure[]{\includegraphics*[angle=0,width=0.45\textwidth]{figs/bfh2gam-xenon.eps}}\\
\subfigure[]{\includegraphics*[angle=0,width=0.45\textwidth]{figs/bfh2ww-xenon.eps}}
\subfigure[]{\includegraphics*[angle=0,width=0.45\textwidth]{figs/bfh2zz-xenon.eps}}
\caption{Effective production rates of a (a) $b\bar b$, (b) $\gamma \gamma$, (c) $W^+W^-$, and (d) $ZZ$ through the heavy Higgs boson versus its mass for the XENON100 $90\%$ C.L. consistent region.  Suppression with respect to the SM expectations are caused by the singlet scalar-Higgs mixing and the competing decay modes $H_2\to H_1 H_1$ and $H_2\to AA$.}
\label{fig:BFh2SMxenon}
\end{center}
\end{figure}

\item Large decay rates of $H_2\to AA$, with $A$ undetected, will give a large missing energy in collider events.  Higgs production via vector-boson fusion (VBF) followed by a Higgs decay to $AA$ may be cleanly extracted from the background of QCD and electroweak $W^\pm,Z+jj$ production~\cite{Eboli:2000ze}.  In addition, it may be possible to observe an invisible Higgs via the $Z$-Higgs-strahlung channel~\cite{Davoudiasl:2004aj}.  With $30$ fb$^{-1}$ and $\sqrt{s}=14$ TeV, ATLAS is expected to probe, at the $5\sigma$ level, the invisible branching fraction of a Higgs boson with SM-strength couplings down to $25\%-30\%$ for the relevant $H_{2}$ mass range~\cite{Schram:2008zz}.  The invisible branching fraction of $H_{2}$ decay in the CSM can be nearly $100\%$, as shown in Fig.~\ref{fig:hcascadecogent}a. Often, this branching is near $50\%$, comparable to the cascade decay discussed below, allowing ATLAS to probe most of the region allowed by CoGeNT.  For XENON100, the invisible branching ratio of $H_2$ can have more of a range, but is also expected to be well covered by the ATLAS analysis, see Fig.~\ref{fig:hcascadexenon}a.

\begin{figure}[htpb]
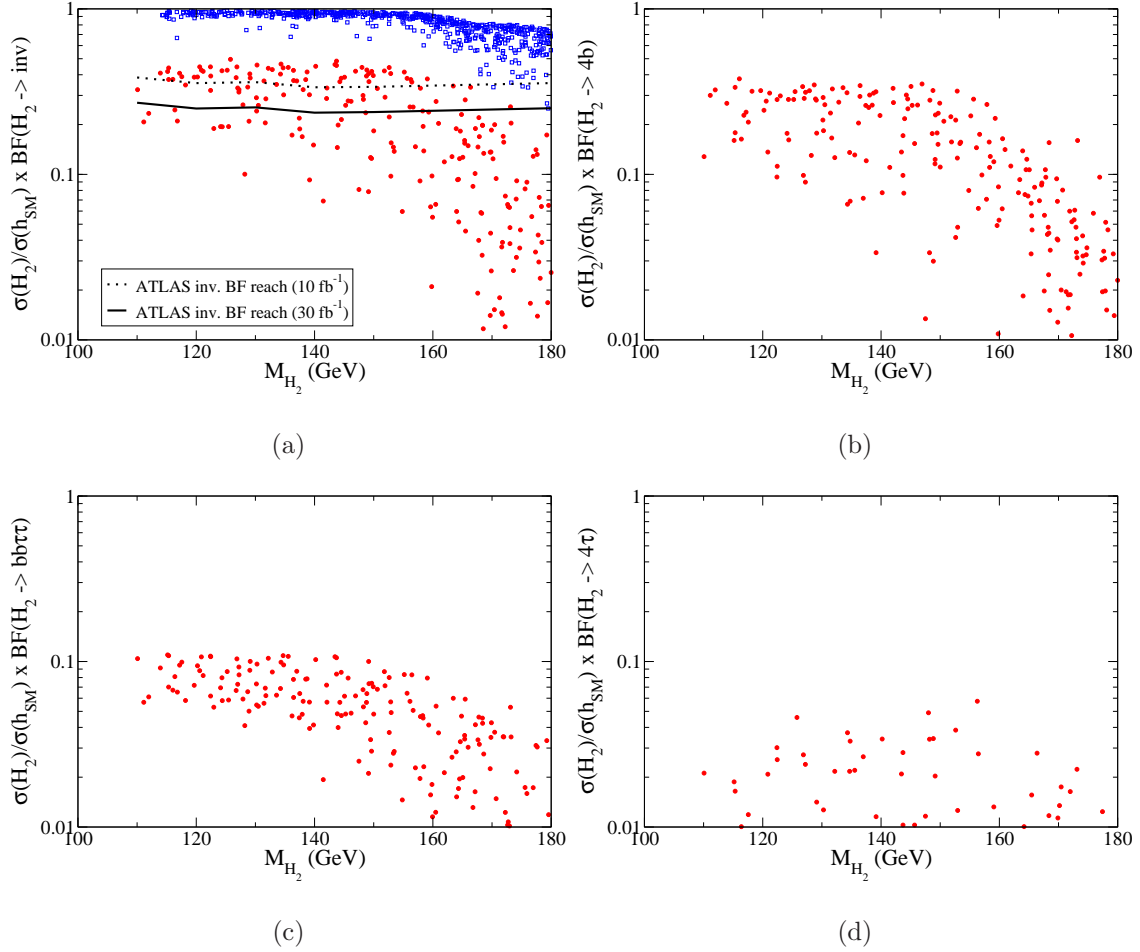

\begin{center}
\subfigure[]{\includegraphics*[angle=0,width=0.45\textwidth]{figs/bfh2toinv-cogent.eps}}
\subfigure[]{\includegraphics*[angle=0,width=0.45\textwidth]{figs/bfh2to4b-cogent.eps}}\\
\subfigure[]{\includegraphics*[angle=0,width=0.45\textwidth]{figs/bfh2tobbtau-cogent.eps}}
\subfigure[]{\includegraphics*[angle=0,width=0.45\textwidth]{figs/bfh2to4tau-cogent.eps}}
\caption{Effective production rates of (a) $\met$, (b) $b\bar b b \bar b$, (c) $b\bar b \tau^+\tau^-$, and (d) $\tau^+\tau^-\tau^+\tau^-$ through the $H_2$ state for the $90\%$ C.L. boundary of the CoGeNT allowed region.}
\label{fig:hcascadecogent}
\end{center}
\end{figure}

\begin{figure}[htpb]
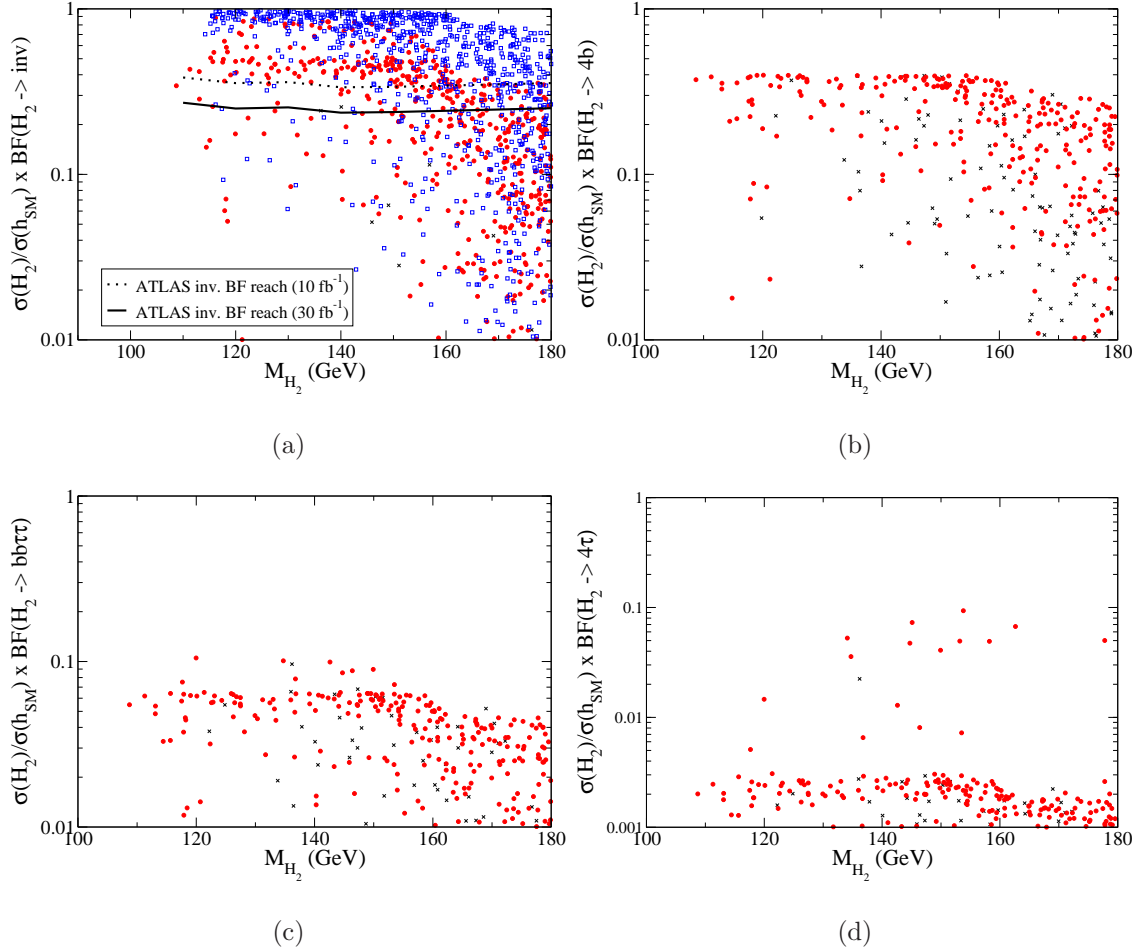

\begin{center}
\subfigure[]{\includegraphics*[angle=0,width=0.45\textwidth]{figs/bfh2toinv-xenon.eps}}
\subfigure[]{\includegraphics*[angle=0,width=0.45\textwidth]{figs/bfh2to4b-xenon.eps}}\\
\subfigure[]{\includegraphics*[angle=0,width=0.45\textwidth]{figs/bfh2tobbtau-xenon.eps}}
\subfigure[]{\includegraphics*[angle=0,width=0.45\textwidth]{figs/bfh2to4tau-xenon.eps}}
\caption{Effective production rates of (a) $\met$, (b) $b\bar b b \bar b$, (c) $b\bar b \tau^+\tau^-$, and (d) $\tau^+\tau^-\tau^+\tau^-$ through the $H_2$ state for the XENON100 $90\%$ C.L. consistent region.}
\label{fig:hcascadexenon}
\end{center}
\end{figure}

\item Large decay rates of $H_{2}\to H_{1}H_{1}$ : This generic class of decays has been studied in the context of the NMSSM~\cite{Dermisek:2005gg,Barger:2006sk,Dermisek:2006wr,Chang:2008cw} and Little Higgs models~\cite{Cheung:2007sva, Cheung:2008zu}.  In addition, model independent analyses of this mode have illustrated the sensitivity of the Tevatron and the LHC experiments to these exotic decay modes~\cite{Carena:2007jk, Cheung:2007sva}.  The effective production rates, ${\sigma(H_2)\over \sigma(h_{SM})}\times\text{BF}(H_2\to 2 H_1 \to 2X+2Y)$, are shown in Figs.~\ref{fig:hcascadecogent} and~\ref{fig:hcascadexenon}.  Once produced, the $H_{1}$ pair will subsequently decay into:
\begin{itemize}
\item $b\bar b + b\bar b$ : This channel may be probed at hadron colliders~\cite{Carena:2007jk}.  At the Tevatron, the production signal is $W^\pm,Z$-Higgs-strahlung, which may yield a few events that are almost background free.  Therefore, detection of this mode is statistics limited.  The LHC also has sensitivity to this channel even though the QCD backgrounds are much larger there.  For the CoGeNT [Fig.~\ref{fig:hcascadecogent}b] and XENON100 [Fig.~\ref{fig:hcascadexenon}b], these branching fractions can  be as large as $40\%$.  This value is just slightly below the benchmark case analyzed in~\cite{Carena:2007jk}, where $k_2 \times \text{BF}(H_{2}\to H_{1}H_{1}) \times \text{BF}(H_{1}\to b\bar b)^{2} = 0.5$, assuming a SM-like $H_{2}$.  In that benchmark scenario, with $M_{H_{2}}=120$ GeV and $M_{H_{1}}=30$ GeV, $5\sigma$ discovery could be achieved at the LHC at $14$ TeV with less than $30$ fb$^{-1}$ of integrated luminosity.  The tagging of soft $b$-quarks is crucial.

\item $b\bar b + \tau^+ \tau^-$ :  This mode may also be probed at the Tevatron.  For the CoGeNT [Fig.~\ref{fig:hcascadecogent}c] and XENON100 [Fig.~\ref{fig:hcascadexenon}c] regions, the branching fractions can be up to $10\%$ which is slightly above the benchmark case analyzed in~\cite{Carena:2007jk}, where $2\times k_2 \times \text{BF}(H_{2}\to H_{1}H_{1}) \times \text{BF}(H_{1}\to b\bar{b}) \times \text{BF}(H_{1}\to \tau^{+}\tau^{-}) = 0.088$, resulting in a rate of about $0.3$ fb.  Therefore, this channel is severely statistics limited. At the LHC, the reducible background is problematic and excellent jet rejection would be needed~\cite{Carena:2007jk}.   

\item $\tau^+ \tau^- + \tau^+ \tau^-$:  This mode can be large when the $H_{1}\to b\bar{b}$ mode is kinematically suppressed.   The CSM branching fractions of this mode for the CoGeNT [Fig.~\ref{fig:hcascadecogent}d] and XENON100 [Fig.~\ref{fig:hcascadexenon}d] regions can be up to about $5\%$.  Searches through the VBF production of $H_{2}$ with semileptonic and hadronic decays of the $\tau$ final states to $\mu^\pm$ and jets can be probed at the LHC~\cite{Adam:2008uu,Forshaw:2007ra}, while the Tevatron may also see an excess in multilepton events with a light $H_{2}$~\cite{Graham:2006tr}.  

\item $b\bar b + \met$ : The branching fraction of this mode is smaller than $1\%$ for both CoGeNT and XENON100.  In addition, the soft $b$-quarks may make it difficult to extract the signal from QCD backgrounds.  Because of the $\met$, full reconstruction of the $H_{2}$ mass is lost using traditional methods.  Therefore, this decay mode does not look promising at hadron colliders.

\item $\tau^+ \tau^- + \met$ : The branching fraction of this mode is smaller than $0.1\%$ for both CoGeNT and XENON100.  This channel yelds soft $\tau$-leptons.  The decays of the $\tau$-leptons yield one additional $\met$ component that will be back to back to the $\met$ from the $H_{1}\to AA$ decay.  Since the $\met$ should be softer than in the $b\bar b+\met$ mode, this decay mode is not promising at hadron colliders.

\item $\met$ : This mode is equivalent to the above $H_{2}\to AA$ invisible decay mode, but each $H_{1}$ state decays to $AA$ and shown as the red circles in Fig.~\ref{fig:hcascadecogent}a for CoGeNT and Fig.~\ref{fig:hcascadexenon}a for XENON100 .  Therefore, the VBF channel for detecting $H$ to $\met$ is applicable.
\end{itemize}
\end{enumerate}

\begin{figure}[htpb]
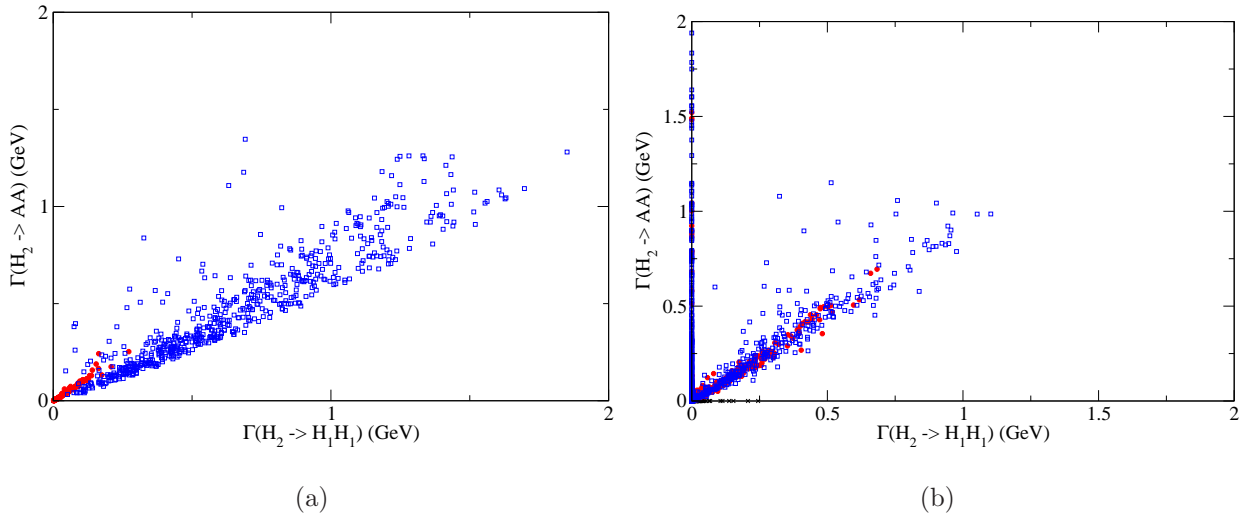

\begin{center}
\subfigure[]{\includegraphics*[angle=0,width=0.49\textwidth]{figs/widthAAvh1h1-cogent.eps}}~
\subfigure[]{\includegraphics*[angle=0,width=0.49\textwidth]{figs/widthAAvh1h1-xenon.eps}}
\caption{The $H_{2}\to AA$ partial width versus the $H_{2}\to H_{1}H_{1}$ partial width for (a)  the $90\%$ C.L. boundary of the CoGeNT allowed region and (b) the XENON100 $90\%$ C.L. consistent region.  Both can simultaneously be large, providing a dual signature of invisible and cascade Higgs decays.}
\label{fig:widaavsh1}
\end{center}
\end{figure}

There is strong competition between the purely invisible $H_{2}\to AA$ and the cascade decay $H_{2}\to H_{1}H_{1}$, presented as the partial widths of these two decay modes in Fig.~\ref{fig:widaavsh1}.  Often, the couplings $g_{H_{2}AA}$ and $g_{H_{2}H_{1}H_{1}}$ are nearly equivalent when there is a large singlet content of the $H_{1}$, providing roughly the same partial widths.  Because of these effects, both large missing energy signatures and Higgs cascade decays can be realized in the same model parameter space when these modes are kinematically accessible.

%%%%%%%%%%%%%%%%%%%%%%%%%%%%%%%%%%
\subsection{Electroweak Phase Transition}
\label{sect:ewpt}
%%%%%%%%%%%%%%%%%%%%%%%%%%%%%%%%%%

The CSM can also provide a strong first order electroweak phase transition (EWPT) in the early universe that is required for electroweak baryogenesis~\cite{Profumo:2007wc,Barger:2008jx}.  In order to prevent the washout of the baryon asymmetry produced during the phase transition, the inequality
\begin{equation}
{v(T_C)\over T_C} \gtrsim 1.
\end{equation}
must be satisfied. Here $T_C$ is the phase transtion critical temperature and $v(T)$ is the $SU(2)_L$ vev at temperature $T$.  It has been shown that to satisfy this inequality, one requires a value of $\delta_2$ that is negative and with a singlet vev  $v_S\lesssim 100$ GeV~\cite{Barger:2008jx}.  

\begin{figure}[htpb]
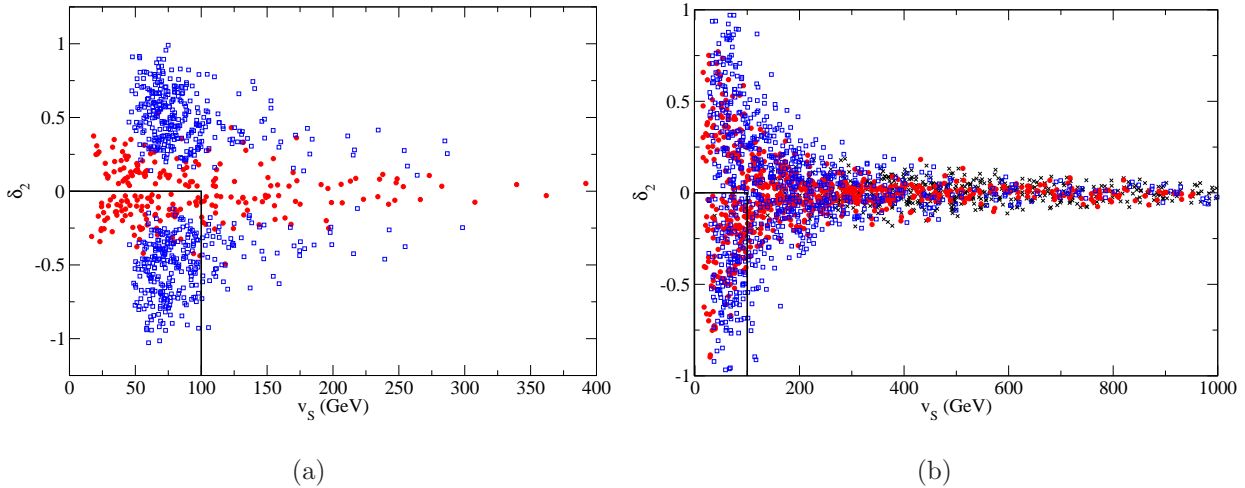

\begin{center}
\subfigure[]{\includegraphics*[angle=0,width=0.49\textwidth]{figs/del2vsvS-cogent.eps}}~
\subfigure[]{\includegraphics*[angle=0,width=0.49\textwidth]{figs/del2vsvS-xenon.eps}}
\caption{Values of $\delta_2$ versus $v_S$ for (a) the $90\%$ C.L. boundary of the CoGeNT allowed region and (b) the XENON100 $90\%$ C.L. consistent region.  The region within the black box can be consistent with an EWPT required for electroweak baryogenesis.}
\label{fig:ewpt}
\end{center}
\end{figure}

In Fig.~\ref{fig:ewpt}, we show the correlation between $\delta_2$ and $v_S$ after all constraints are applied.  The model can naturally provide an EWPT regardless of whether we restrict ourselves to the CoGeNT or XENON100 regions.  Because of the CoGeNT preference for low DM mass, the singlet vev is required to be relatively small, yielding EWPT more easily.  We focus on the region satisfying the EWPT requirements, which is within the black box.  We find that the sign of $\delta_2$ has little effect on many observables such as relic density and Higgs production.   The effect of flipping this sign may be noticeable in the interference terms of $AA\to H_i H_j$~\cite{Barger:2008jx}.  Establishing EWPT in this model requires measurement of the couplings in the scalar potential in Eq.~\ref{eq:potential} that give rise to the first order phase transition.

%%%%%%%%%%%%%%%%%%%%%%%%%%%%%%%%%%
\section{Conclusions}
\label{sect:conclusions}
%%%%%%%%%%%%%%%%%%%%%%%%%%%%%%%%%%

We have explored the viability of the complex scalar singlet model (CSM) as an explanation of the CoGeNT and DAMA/LIBRA DM signals without channeling effects and the nonexcluded XENON100 region. The model has a DM particle that has Higgs-only interactions with the SM fields.  There is also an additional real scalar field that mixes with the SM-Higgs field.  The light Higgs boson must be dominantly singlet to reproduce the WMAP relic density measurement with thermal DM production in the early Universe. The DM particle can have a scattering cross section that is within the CoGeNT and XENON100 regions and satisfies the relic abundance observed by WMAP, the LEP limits on the $ZZH_{i}$ mixing, and on the cascade decays of the $H_2$ to $H_1$ pairs to $b\bar b + b\bar b$, $b\bar b + \tau^+\tau^-$, and $\tau^+\tau^-+\tau^+\tau^-$, the OPAL limit on generic Higgs production, and EWPO constraints. 

For the CoGeNT, DAMA/LIBRA, and XENON100 allowed regions, we find
\begin{itemize}

\item The complex singlet extended standard model has the attractive feature that the DM mass can range from a few GeV to a few TeV with a thermal relic density as observed by WMAP.  It can also be consistent with experimental constraints on a low mass Higgs boson.

\item Given a mass of $A$ of about $10$ GeV from the putative CoGeNT DM signal, the $H_{1}$ mass is predicted to lie in the ranges $9$ to $15$ GeV or $30$ to $70$ GeV, depending on which side of the $H_1$ annihilation resonance the $A$ state lies.  We also find the $H_{2}$ has a mass between $110$ to $180$ GeV with the lower and upper bound arising from the LEP $ZZH_{2}$ and EWPO constraints, respectively.  To also be consistent with the CoGeNT and DAMA/LIBRA DM signal, the DM mass is restricted to the $\sim7$ GeV region, for which the SI cross section is $\sim10^{-40}\text{ cm}^2$~\cite{Hooper:2010uy}.  In our scans, we find only a few points that fall within the DAMA/LIBRA allowed region.

\item Since $A$ is stable, the decays of $H_{2}$ to $AA$ result in large missing energy.  This invisible decay mode can be probed at the LHC via weak boson fusion and provide a robust test of the model.   Alternatively, the $H_{2}$ can decay to $WW$ and $ZZ$ with sufficiently large branching fractions as to allow their detection at the LHC.

\item The decays of $H_{2}$ to $H_{1}H_{1}$, followed by the decays of  $H_{1}$ to SM particles, can be probed at the LHC through the $4b$, $4\tau$, and $2b+2\tau$ channels.  Alternatively, when the $H_{1}$ decays to $AA$, it contributes to the missing energy.  See Figs.~\ref{fig:mh1vsma},~\ref{fig:BFh2SMcogent},~\ref{fig:BFh2SMxenon},~\ref{fig:hcascadecogent} and~\ref{fig:hcascadexenon}.

\item The lightest Higgs state may be lighter than the DM state, resulting in the annihilation mode $AA\to H_{1}H_{1}$.  Because of the low mass of $H_{1}$, the decay to $b\bar{b}$ may be kinematically suppressed, resulting in an enhanced branching to $\tau^{+}\tau^{-}$.  This prediction can be tested by comparing the $H_{2}\to 2b+2\tau$ and $H_{2}\to 4\tau$ branching fractions.

\item The SM-like $H_{2}$ state can often decay to $AA$ and to $H_{1}H_{1}$, with comparable partial widths of the two modes, providing a dual missing energy and cascade decay signature at the LHC.  See Fig.~\ref{fig:widaavsh1}.

\end{itemize}

In summary, the complex scalar singlet model is an attractive explanation of the CoGeNT candidate DM signal in that the model naturally accommodates light DM, the CoGeNT DM scattering cross section, and the WMAP DM relic density.  A spin-independent cross section of $10^{-40}\text{ cm}^2$ is consistent with both CoGeNT and DAMA/LIBRA data for a DM mass of $\sim7$ GeV.  Moreover, the model is testable at the LHC, especially via the $qq^\prime\to qq^\prime H_2 \to qq^\prime \met$ channel or via the cascade decay of the $H_2$ to multiple $b$-jets or $\tau$-leptons in the $q\bar q \to W/Z + H_2$ channel.  Such decay modes can dominate the standard $h_{SM}\to WW/ZZ$ channels.  Even if the CoGeNT and DAMA/LIBRA signals should not be confirmed and the XENON100 exclusion holds, the model remains viable, but its predictions for Higgs phenomenology become less specific.

%%%%%%%%%%%%%%%%%%%%%%%%%%%%%%%%%%
\section{Acknowledgments}
%%%%%%%%%%%%%%%%%%%%%%%%%%%%%%%%%%

We thank Juan Collar, Patrick Draper, Wai-Yee Keung, Jason Kumar, D. Hooper, Ian Low, Danny Marfatia, Uwe Oberlack, and Kathryn Zurek for helpful interactions and we thank Paul Langacker and Michael Ramsey-Musolf for earlier collaboration on the CSM.  This work was supported in part by the U.S. Department of Energy Division of High Energy Physics under Grants No. DE-FG02-95ER40896, No. DE-FG02-05ER41361, No. DE-FG02-08ER41531, No. DE-FG02-91ER40684 and Contract No. DE-AC02-06CH11357, by the Wisconsin Alumni Research Foundation, and by the National Science Foundation Grant No. PHY-0503584.

%%%%%%%%%%%%%%%%%%%%%%%%%%%%%%%%%%
\bibliographystyle{apsrev}
\bibliography{cogent}

\begin{thebibliography}{10}

\bibitem{Larson:2010gs}
D.~Larson {\em et~al.},
\newblock (2010), arXiv:1001.4635.
%%CITATION = 1001.4635;%%

\bibitem{Adriani:2008zr}
O.~Adriani {\em et~al.} (PAMELA Collaboration),
\newblock Nature {\bf 458}, 607 (2009).
%%CITATION = 0810.4995;%%

\bibitem{Abdo:2009zk}
A.~A. Abdo {\em et~al.} (The Fermi LAT Collaboration),
\newblock Phys. Rev. Lett. {\bf 102}, 181101 (2009).
%%CITATION = 0905.0025;%%

\bibitem{Aharonian:2009ah}
F. Aharonian (H.E.S.S. Collaboration),
\newblock Astron. Astrophys. {\bf 508}, 561 (2009), arXiv:0905.0105.
%%CITATION = 0905.0105;%%

\bibitem{Abdo:2010nc}
A.~A. Abdo {\em et~al.},
\newblock Phys. Rev. Lett. {\bf 104}, 091302 (2010), arXiv:1001.4836.
%%CITATION = 1001.4836;%%

\bibitem{Barger:2001ur}
V.~D. Barger, F.~Halzen, D.~Hooper, and C.~Kao,
\newblock Phys. Rev. {\bf D65}, 075022 (2002), arXiv:hep-ph/0105182.
%%CITATION = HEP-PH/0105182;%%

\bibitem{Cirelli:2005gh}
M.~Cirelli {\em et~al.},
\newblock Nucl. Phys. {\bf B727}, 99 (2005), arXiv:hep-ph/0506298.
%%CITATION = HEP-PH/0506298;%%

\bibitem{GonzalezGarcia:2005xw}
M.~C. Gonzalez-Garcia, F.~Halzen, and M.~Maltoni,
\newblock Phys. Rev. {\bf D71}, 093010 (2005), arXiv:hep-ph/0502223.
%%CITATION = HEP-PH/0502223;%%

\bibitem{Barger:2007xf}
V.~Barger, W.-Y. Keung, G.~Shaughnessy, and A.~Tregre,
\newblock Phys. Rev. {\bf D76}, 095008 (2007), arXiv:0708.1325.
%%CITATION = 0708.1325;%%

\bibitem{Barger:2007hj}
V.~Barger, W.-Y. Keung, and G.~Shaughnessy,
\newblock Phys. Lett. {\bf B664}, 190 (2008), arXiv:0709.3301.
%%CITATION = 0709.3301;%%

\bibitem{Ahmed:2009zw}
Z.~Ahmed {\em et~al.} (The CDMS-II Collaboration),
\newblock (2009), arXiv:0912.3592.
%%CITATION = 0912.3592;%%

\bibitem{Bottino:2009km}
A.~Bottino, F.~Donato, N.~Fornengo, and S.~Scopel,
\newblock Phys. Rev. {\bf D81}, 107302 (2010).
%%CITATION = 0912.4025;%%

\bibitem{Cheung:2009wb}
K.~Cheung and T.-C. Yuan,
\newblock Phys. Lett. {\bf B685}, 182 (2010).
%%CITATION = 0912.4599;%%

\bibitem{He:2009yd}
X.-G. He, T.~Li, X.-Q. Li, J.~Tandean, and H.-C. Tsai,
\newblock Phys. Lett. {\bf B688}, 332 (2010).
%%CITATION = 0912.4722;%%

\bibitem{He:2010nt}
X.-G. He, S.-Y. Ho, J.~Tandean, and H.-C. Tsai,
\newblock (2010), arXiv:1004.3464.
%%CITATION = 1004.3464;%%

\bibitem{Bernabei:2008yi}
R.~Bernabei {\em et~al.} (DAMA Collaboration),
\newblock Eur. Phys. J. {\bf C56}, 333 (2008).
%%CITATION = 0804.2741;%%

\bibitem{Bottino:2008mf}
A.~Bottino, F.~Donato, N.~Fornengo, and S.~Scopel,
\newblock Phys. Rev. {\bf D78}, 083520 (2008).
%%CITATION = 0806.4099;%%

\bibitem{Bernabei:2010mq}
R.~Bernabei {\em et~al.},
\newblock (2010), arXiv:1002.1028.
%%CITATION = 1002.1028;%%

\bibitem{Petriello:2008jj}
F.~Petriello and K.~M. Zurek,
\newblock JHEP {\bf 09}, 047 (2008).
%%CITATION = 0806.3989;%%

\bibitem{Savage:2008er}
C.~Savage, G.~Gelmini, P.~Gondolo, and K.~Freese,
\newblock JCAP {\bf 0904}, 010 (2009).
%%CITATION = 0808.3607;%%

\bibitem{Feldstein:2009np}
B.~Feldstein, A.~L. Fitzpatrick, E.~Katz, and B.~Tweedie,
\newblock JCAP {\bf 1003}, 029 (2010).
%%CITATION = 0910.0007;%%

\bibitem{Bozorgnia:2010xy}
N.~Bozorgnia, G.~B. Gelmini, and P.~Gondolo,
\newblock (2010), arXiv:1006.3110.
%%CITATION = 1006.3110;%%

\bibitem{Aalseth:2010vx}
C.~E. Aalseth {\em et~al.} (CoGeNT Collaboration),
\newblock (2010), arXiv:1002.4703.
%%CITATION = 1002.4703;%%

\bibitem{Essig:2010ye}
R.~Essig, J.~Kaplan, P.~Schuster, and N.~Toro,
\newblock (2010), arXiv:1004.0691.
%%CITATION = 1004.0691;%%

\bibitem{Graham:2010ca}
P.~W. Graham, R.~Harnik, S.~Rajendran, and P.~Saraswat,
\newblock (2010), arXiv:1004.0937.
%%CITATION = 1004.0937;%%

\bibitem{Andreas:2008xy}
S.~Andreas, T.~Hambye, and M.~H.~G. Tytgat,
\newblock JCAP {\bf 0810}, 034 (2008).
%%CITATION = 0808.0255;%%

\bibitem{Chang:2010yk}
S.~Chang, J.~Liu, A.~Pierce, N.~Weiner, and I.~Yavin,
\newblock (2010), arXiv:1004.0697.
%%CITATION = 1004.0697;%%

\bibitem{Andreas:2010dz}
S.~Andreas, C.~Arina, T.~Hambye, F.-S. Ling, and M.~H.~G. Tytgat,
\newblock (2010), arXiv:1003.2595.
%%CITATION = 1003.2595;%%

\bibitem{Fitzpatrick:2010em}
A.~L. Fitzpatrick, D.~Hooper, and K.~M. Zurek,
\newblock (2010), arXiv:1003.0014.
%%CITATION = 1003.0014;%%

\bibitem{Kaplan:1992db}
D.~B. Kaplan,
\newblock Phys. Rev. Lett. {\bf 68}, 741 (1992).

\bibitem{Hooper:2004dc}
D.~Hooper, J.~March-Russell, and S.~M. West,
\newblock Phys. Lett. {\bf B605}, 228 (2005).
%%CITATION = HEP-PH/0410114;%%

\bibitem{Farrar:2005zd}
G.~R. Farrar and G.~Zaharijas,
\newblock Phys. Rev. Lett. {\bf 96}, 041302 (2006).
%%CITATION = HEP-PH/0510079;%%

\bibitem{Kitano:2008tk}
R.~Kitano, H.~Murayama, and M.~Ratz,
\newblock Phys. Lett. {\bf B669}, 145 (2008).
%%CITATION = 0807.4313;%%

\bibitem{Kaplan:2009ag}
D.~E. Kaplan, M.~A. Luty, and K.~M. Zurek,
\newblock Phys. Rev. {\bf D79}, 115016 (2009).
%%CITATION = 0901.4117;%%

\bibitem{Cohen:2010kn}
T.~Cohen, D.~J. Phalen, A.~Pierce, and K.~M. Zurek,
\newblock (2010), arXiv:1005.1655.
%%CITATION = 1005.1655;%%

\bibitem{Bottino:2003cz}
A.~Bottino, F.~Donato, N.~Fornengo, and S.~Scopel,
\newblock Phys. Rev. {\bf D69}, 037302 (2004).
%%CITATION = HEP-PH/0307303;%%

\bibitem{Bottino:2003iu}
A.~Bottino, F.~Donato, N.~Fornengo, and S.~Scopel,
\newblock Phys. Rev. {\bf D68}, 043506 (2003).
%%CITATION = HEP-PH/0304080;%%

\bibitem{Feldman:2010ke}
D.~Feldman, Z.~Liu, and P.~Nath,
\newblock (2010), arXiv:1003.0437.
%%CITATION = 1003.0437;%%

\bibitem{Kuflik:2010ah}
E.~Kuflik, A.~Pierce, and K.~M. Zurek,
\newblock (2010), arXiv:1003.0682.
%%CITATION = 1003.0682;%%

\bibitem{Dermisek:2005gg}
R.~Dermisek and J.~F. Gunion,
\newblock Phys. Rev. {\bf D73}, 111701 (2006).
%%CITATION = HEP-PH/0510322;%%

\bibitem{Ellwanger:2004gz}
U.~Ellwanger, J.~F. Gunion, C.~Hugonie, and S.~Moretti,
\newblock (2004), arXiv:hep-ph/0401228.
%%CITATION = HEP-PH/0401228;%%

\bibitem{Accomando:2006ga}
E.~Accomando {\em et~al.},
\newblock (2006), arXiv:hep-ph/0608079.
%%CITATION = HEP-PH/0608079;%%

\bibitem{Barger:2006sk}
V.~Barger, P.~Langacker, and G.~Shaughnessy,
\newblock Phys. Rev. {\bf D75}, 055013 (2007).
%%CITATION = HEP-PH/0611239;%%

\bibitem{Cerdeno:2008ep}
D.~G. Cerdeno, C.~Munoz, and O.~Seto,
\newblock Phys. Rev. {\bf D79}, 023510 (2009).
%%CITATION = 0807.3029;%%

\bibitem{Cerdeno:2009dv}
D.~G. Cerdeno and O.~Seto,
\newblock JCAP {\bf 0908}, 032 (2009).
%%CITATION = 0903.4677;%%

\bibitem{Aprile:2010um}
E.~Aprile {\em et~al.} (XENON100 Collaboration),
\newblock (2010), arXiv:1005.0380.
%%CITATION = 1005.0380;%%

\bibitem{Collaboration:2010er}
E.~Aprile {\em et~al.} (XENON100 Collaboration),
\newblock (2010), arXiv:1005.2615.
%%CITATION = 1005.2615;%%

\bibitem{Sorensen:2010hq}
P.~Sorensen,
\newblock (2010), arXiv:1007.3549.
%%CITATION = 1007.3549;%%

\bibitem{Collar:2010gg}
J.~I. Collar and D.~N. McKinsey,
\newblock (2010), arXiv:1005.0838.
%%CITATION = 1005.0838;%%

\bibitem{Collar:2010gd}
J.~I. Collar and D.~N. McKinsey,
\newblock (2010), arXiv:1005.3723.
%%CITATION = 1005.3723;%%

\bibitem{Barger:2008jx}
V.~Barger, P.~Langacker, M.~McCaskey, M.~Ramsey-Musolf, and G.~Shaughnessy,
\newblock Phys. Rev. {\bf D79}, 015018 (2009).
%%CITATION = 0811.0393;%%

\bibitem{McDonald:1993ex}
J.~McDonald,
\newblock Phys. Rev. {\bf D50}, 3637 (1994).
%%CITATION = HEP-PH/0702143;%%

\bibitem{Burgess:2000yq}
C.~P. Burgess, M.~Pospelov, and T.~ter Veldhuis,
\newblock Nucl. Phys. {\bf B619}, 709 (2001).
%%CITATION = HEP-PH/0011335;%%

\bibitem{BahatTreidel:2006kx}
O.~Bahat-Treidel, Y.~Grossman, and Y.~Rozen,
\newblock JHEP {\bf 05}, 022 (2007).
%%CITATION = HEP-PH/0611162;%%

\bibitem{OConnell:2006wi}
D.~O'Connell, M.~J. Ramsey-Musolf, and M.~B. Wise,
\newblock Phys. Rev. {\bf D75}, 037701 (2007).
%%CITATION = HEP-PH/0611014;%%

\bibitem{Barger:2007im}
V.~Barger, P.~Langacker, M.~McCaskey, M.~J. Ramsey-Musolf, and G.~Shaughnessy,
\newblock Phys. Rev. {\bf D77}, 035005 (2008).
%%CITATION = 0706.4311;%%

\bibitem{Asano:2010yi}
M.~Asano and R.~Kitano,
\newblock Phys. Rev. {\bf D81}, 054506 (2010).
%%CITATION = 1001.0486;%%

\bibitem{Arina:2010an}
C.~Arina, F.-X. Josse-Michaux, and N.~Sahu,
\newblock (2010), arXiv:1004.3953.
%%CITATION = 1004.3953;%%

\bibitem{Bandyopadhyay:2010cc}
A.~Bandyopadhyay, S.~Chakraborty, A.~Ghosal, and D.~Majumdar,
\newblock (2010), arXiv:1003.0809.
%%CITATION = 1003.0809;%%

\bibitem{Patt:2006fw}
B.~Patt and F.~Wilczek,
\newblock (2006), arXiv:hep-ph/0605188.
%%CITATION = HEP-PH/0605188;%%

\bibitem{Gelmini:2006pw}
G.~B. Gelmini and P.~Gondolo,
\newblock Phys. Rev. {\bf D74}, 023510 (2006).
%%CITATION = HEP-PH/0602230;%%

\bibitem{Salati:2002md}
P.~Salati,
\newblock Phys. Lett. {\bf B571}, 121 (2003).
%%CITATION = ASTRO-PH/0207396;%%

\bibitem{Profumo:2003hq}
S.~Profumo and P.~Ullio,
\newblock JCAP {\bf 0311}, 006 (2003).
%%CITATION = HEP-PH/0309220;%%

\bibitem{Rosati:2003yw}
F.~Rosati,
\newblock Phys. Lett. {\bf B570}, 5 (2003).
%%CITATION = HEP-PH/0302159;%%

\bibitem{Chung:2007vz}
D.~J.~H. Chung, L.~L. Everett, and K.~T. Matchev,
\newblock Phys. Rev. {\bf D76}, 103530 (2007).
%%CITATION = 0704.3285;%%

\bibitem{Belanger:2008sj}
G.~Belanger, F.~Boudjema, A.~Pukhov, and A.~Semenov,
\newblock Comput. Phys. Commun. {\bf 180}, 747 (2009).
%%CITATION = 0803.2360;%%

\bibitem{Belanger:2010gh}
G.~Belanger {\em et~al.},
\newblock (2010), arXiv:1004.1092.
%%CITATION = 1004.1092;%%

\bibitem{Abbiendi:2002qp}
G.~Abbiendi {\em et~al.} (OPAL Collaboration),
\newblock Eur. Phys. J. {\bf C27}, 311 (2003).
%%CITATION = HEP-EX/0206022;%%

\bibitem{Sopczak:2005mc}
A.~Sopczak,
\newblock (2005), arXiv:hep-ph/0502002.
%%CITATION = HEP-PH/0502002;%%

\bibitem{Abdallah:2003ry}
J.~Abdallah {\em et~al.} (DELPHI Collaboration),
\newblock Eur. Phys. J. {\bf C32}, 475 (2004).
%%CITATION = HEP-EX/0401022;%%

\bibitem{LEP:2001xz}
LEP Higgs Working Group,
\newblock (2001), arXiv:hep-ex/0107032.
%%CITATION = HEP-EX/0107032;%%

\bibitem{Dermisek:2006wr}
R.~Dermisek and J.~F. Gunion,
\newblock Phys. Rev. {\bf D75}, 075019 (2007).
%%CITATION = HEP-PH/0611142;%%

\bibitem{Carena:2007jk}
M.~Carena, T.~Han, G.-Y. Huang, and C.~E.~M. Wagner,
\newblock JHEP {\bf 04}, 092 (2008).
%%CITATION = 0712.2466;%%

\bibitem{Cheung:2007sva}
K.~Cheung, J.~Song, and Q.-S. Yan,
\newblock Phys. Rev. Lett. {\bf 99}, 031801 (2007).
%%CITATION = HEP-PH/0703149;%%

\bibitem{Chang:2008cw}
S.~Chang, R.~Dermisek, J.~F. Gunion, and N.~Weiner,
\newblock Ann. Rev. Nucl. Part. Sci. {\bf 58}, 75 (2008).
%%CITATION = 0801.4554;%%

\bibitem{Aleph:2010aw}
S.~Schael {\em et~al.} (ALEPH Collaboration),
\newblock (2010), arXiv:1003.0705.
%%CITATION = 1003.0705;%%

\bibitem{Amsler20081}
C.~Amsler {\em et~al.},
\newblock Physics Letters B {\bf 667}, 1  (2008),
\newblock Review of Particle Physics.

\bibitem{Ellis:2000ds}
J.~R. Ellis, A.~Ferstl, and K.~A. Olive,
\newblock Phys. Lett. {\bf B481}, 304 (2000).
%%CITATION = HEP-PH/0001005;%%

\bibitem{Barger:2008qd}
V.~Barger, W.-Y. Keung, and G.~Shaughnessy,
\newblock Phys. Rev. {\bf D78}, 056007 (2008).
%%CITATION = 0806.1962;%%

\bibitem{Aaltonen:2010sv}
CDF, T.~Aaltonen {\em et~al.},
\newblock (2010), arXiv:1005.3216.
%%CITATION = 1005.3216;%%

\bibitem{Eboli:2000ze}
O.~J.~P. Eboli and D.~Zeppenfeld,
\newblock Phys. Lett. {\bf B495}, 147 (2000).
%%CITATION = HEP-PH/0009158;%%

\bibitem{Davoudiasl:2004aj}
H.~Davoudiasl, T.~Han, and H.~E. Logan,
\newblock Phys. Rev. {\bf D71}, 115007 (2005).
%%CITATION = HEP-PH/0412269;%%

\bibitem{Schram:2008zz}
M.~Schram,
\newblock PROQUEST-1563019011.

\bibitem{Cheung:2008zu}
K.~Cheung, J.~Song, P.~Tseng, and Q.-S. Yan,
\newblock Phys. Rev. {\bf D78}, 055015 (2008).
%%CITATION = 0806.4411;%%

\bibitem{Adam:2008uu}
N.~E. Adam {\em et~al.},
\newblock (2008), arXiv:0803.1154.
%%CITATION = 0803.1154;%%

\bibitem{Forshaw:2007ra}
J.~R. Forshaw, J.~F. Gunion, L.~Hodgkinson, A.~Papaefstathiou, and A.~D.
  Pilkington,
\newblock JHEP {\bf 04}, 090 (2008).
%%CITATION = 0712.3510;%%

\bibitem{Graham:2006tr}
P.~W. Graham, A.~Pierce, and J.~G. Wacker,
\newblock (2006), arXiv:hep-ph/0605162.
%%CITATION = HEP-PH/0605162;%%

\bibitem{Profumo:2007wc}
S.~Profumo, M.~J. Ramsey-Musolf, and G.~Shaughnessy,
\newblock JHEP {\bf 08}, 010 (2007).
%%CITATION = 0705.2425;%%

\bibitem{Hooper:2010uy}
D.~Hooper, J.~I. Collar, J.~Hall, and D.~McKinsey,
\newblock (2010), arXiv:1007.1005.
%%CITATION = 1007.1005;%%

\end{thebibliography}
%%%%%%%%%%%%%%%%%%%%%%%%%%%%%%%%%%

\newpage

\end{document}